\documentclass[manuscript,screen]{acmart}

\usepackage{enumitem} 
\usepackage{hyperref}
\usepackage{xurl} 

\usepackage[normalem]{ulem} 
\usepackage{multirow} 

\usepackage{graphicx}   
\usepackage{subcaption} 

\usepackage{pifont}
\newcommand{\cmark}{\ding{51}}%

\usepackage[most]{tcolorbox} 
\usepackage{amsmath}    
\usepackage{xcolor}     
\usepackage[table]{xcolor} 
\usepackage{subcaption}
\usepackage[ruled,vlined,linesnumbered]{algorithm2e} 
\usepackage{comment}
\usepackage[normalem]{ulem}

\newcommand{\re}[1]{#1}

\newtcbox{\mytag}{nobeforeafter,colframe=mycolor,colback=mycolor!30!white,boxrule=0.7pt,arc=0pt,
 boxsep=-3pt,left=6pt,right=6pt,top=4pt,bottom=5pt,tcbox raise base}
\definecolor{mycolor}{rgb}{0.122, 0.435, 0.698}

\newtcolorbox{quotebox}{colback=white,breakable,boxrule=0.4pt,colframe=black,fonttitle=\bfseries,top=1pt,bottom=1pt,left=1pt,right=1pt, before skip=10pt, after skip=10pt} 

\newtcolorbox{implicationbox}{boxrule=0.4pt,colframe=black,fonttitle=\bfseries,top=1pt,bottom=1pt,left=1pt,right=1pt, before skip=5pt, after skip=5pt}

\newtcolorbox{answerbox}{boxrule=0.5pt,arc=3pt, top=1pt,bottom=1pt,left=1pt,right=1pt, before skip=10pt, after skip=10pt}

\makeatletter
\tcbset{
    myhbox/.style 2 args={%
        enhanced, 
        breakable,
        colback=cyan!5,
        colframe=teal!50!black,
        attach boxed title to top left={yshift*=-\tcboxedtitleheight}, 
        title={#2},
        boxed title size=title, 
        boxed title style={%
            sharp corners, 
            rounded corners=northwest, 
            colback=tcbcolframe, 
            boxrule=0pt,
            coltitle=white,
            top=0.5mm, bottom=0.5mm, 
        },
        underlay boxed title={%
            \path[fill=tcbcolframe] (title.south west)--(title.south east) 
                to[out=0, in=180] ([xshift=3mm]title.east)--
                (title.center-|frame.east)
                [rounded corners=\kvtcb@arc] |- 
                (frame.north) -| cycle; 
        },
        left=2mm, right=2mm, top=1mm, bottom=1mm, 
        boxsep=1mm, 
        boxrule=0.6pt, 
        #1
    }
}   
\makeatother

\AtBeginDocument{%
  }

\setcopyright{acmlicensed}
\copyrightyear{2018}
\acmYear{2018}
\acmDOI{XXXXXXX.XXXXXXX}
\acmConference[Conference acronym 'XX]{Make sure to enter the correct
  conference title from your rights confirmation email}{June 03--05,
  2018}{Woodstock, NY}
\acmISBN{978-1-4503-XXXX-X/2018/06}

\begin{document}

\title{When to Use Which? Benchmarking Optimisers for Configurable Systems under Varying Budgets}


\author{Chao Jiang}
\email{cxj249@student.bham.ac.uk}
\affiliation{
	\institution{University of Birmingham}
	\country{United Kingdom}
}

\author{Yulong Ye}
\email{yxy382@student.bham.ac.uk}
\affiliation{
	\institution{IDEAS Lab, University of Birmingham}
	\country{United Kingdom}
}

\author{Tao Chen}
\affiliation{%
  \institution{IDEAS Lab, University of Birmingham}
  \country{United Kingdom}
 }
\email{t.chen@bham.ac.uk}

\author{Miqing Li}
\authornote{Corresponding author: Miqing Li, m.li.8@bham.ac.uk.}
\email{m.li.8@bham.ac.uk}

\affiliation{
	\institution{University of Birmingham}
	\country{United Kingdom}
}

\renewcommand{\shortauthors}{Jiang et al.}

\begin{abstract}

Software configuration tuning is crucial for optimising system performance, and various optimisers have emerged over the last decade. Yet, the time required during the tuning process may vary across systems. In some systems (e.g., \textsc{PostgreSQL}), it may take a few minutes to measure a configuration, whereas in some others (e.g., \textsc{MariaDB}), it can take several hours. Moreover, even within the same system, users may have varying budgets and preferred settings. This naturally raises a question---Given a budget level, which optimiser is the best choice for SE practitioners? This matters because optimisers usually have their own ``comfort zone'' and may perform very differently under distinct budgets. 

In this paper, we aim to answer this question. We systematically evaluate eight well-established optimisers across 22 configurable systems under varying budget levels. We find that, unsurprisingly, model-based optimisers (e.g., SMAC) are well-suited under tight budgets, and model-free optimisers (e.g., GAs) become superior with more generous budgets. However, interestingly, there is one optimiser, FLASH, that performs consistently well on most systems regardless of budgets. We lastly investigate the reasons behind this phenomenon and find that many systems possess good local optima (with large basins of attraction), allowing greedy optimisers (e.g., FLASH) to achieve strong performance.

Source code, data, and supplementary materials of this work are available at~\url{https://anonymous.4open.science/r/Config-W2W-98B2}.

\end{abstract}






\received{20 February 2007}
\received[revised]{12 March 2009}
\received[accepted]{5 June 2009}

\maketitle

\section{Introduction}

Over the past decade, software configuration tuning has been an essential way to optimise a performance concern, e.g., runtime, throughput, or accuracy~\cite{herodotou2020survey,zhao2023automatic,pereira2021learning}. 
However, the tuning process of systems is usually computationally expensive~\cite{jamshidi2016uncertainty,chen2021multi,chen2025accuracy}.
In some systems (e.g., \textsc{PostgreSQL}~\cite{postgresql}), it may take a couple of minutes to measure a configuration~\cite{geng2024emit}, whereas in some other systems (e.g., \textsc{MariaDB}~\cite{mariadb}), it can take several hours~\cite{chen2024adapting}. 
The former may allow a few thousand configurations to be trialled~\cite{zhang2023efficient,kanellis2020too}, while the latter needs to be restricted to at most a few hundred~\cite {chen2024adapting}. 
Moreover, even for the same system, the budget considered can differ. 
Users may have varying budgets or even make ad hoc choices.
For example, when tuning the system \textsc{Storm}~\cite{apache_storm}, some studies consider 400 evaluations~\cite{chen2024adapting}, some 250~\cite{jamshidi2016uncertainty}, and some even down to only 50 evaluations~\cite{nair2018finding}. 

It is known that different optimisers typically have their own ``comfort zone'' and may perform differently under different budgets. One may believe some model-based optimisers like Bayesian Optimisation (BO) work well under a tight budget~\cite{iqbal2023cameo,chen2025accuracy,jiang2025multi}, while some model-free optimisers like Genetic Algorithms (GAs) may be favoured when the budget is more generous~\cite{lopez2016irace,chen2023weights}. 
However, this may not always be the case~\cite{lustosa2024learning}. 
For example, when tuning the CPU load for \textsc{MariaDB}, GAs have been found to outperform model-based optimisers (e.g., BO~\cite{chen2021efficient}) at a budget of 200 evaluations~\cite{chen2024adapting}. Yet, when the budget increases to 400 evaluations, the advantage shifts, with model-based optimisers yielding superior results~\cite{chen2024adapting}.

Given the above, a natural question arises: Among various popular optimisers for software configuration tuning, which is the best choice for SE practitioners under different budget levels? 
In this paper, we attempt to answer this question. Specifically, we first consider the tight budgets commonly used in software configuration tuning (i.e., a few hundred evaluations). Then, we change the setup a bit to see what happens when more generous budgets are available (up to ten thousand evaluations, for instance, considered in~\cite{acher2019learning,oh2023finding}). 
In addition, we analyse and discuss the reasons behind the observed results through fitness landscape analysis, which facilitates an intuitive grasp of complex search spaces for understanding how the configuration landscape of systems influences tuning quality~\cite{vskvorc2020understanding}.
\re{It is worth noting that this work considers configurable software systems under unknown-constraint settings, where constraint violations are observed \emph{a posteriori}, i.e., a configuration is identified as invalid only after it has been executed or measured. This is because explicit constraint information (e.g., feature models) is not available for most of the systems considered in our study, such as \textsc{Storm} and \textsc{SQLite}. Scenarios in which constraints can be handled \emph{a priori}, i.e., invalid configurations can be identified before execution or measurement, are outside the scope of this study, e.g., Software Product Line (SPL) optimisation~\cite{oh2023finding,hierons2016sip,oh2017finding,guegain2023configuration}.}

To summarise, we empirically consider several representative optimisers across a wide range of configurable systems under different budgets. 
Specifically,

\begin{itemize}
    \item We systematically evaluate eight model-based and model-free optimisers on 22 configurable software systems under varying evaluation budgets (up to 10,000 evaluations). 

    \item We further explain the observed phenomena through fitness landscape analysis metrics on software systems.

    
    \item We provide guidance for practitioners on choosing suitable optimisers under different budget levels, and provide recommendations for configuration tuning researchers to design new optimisers. 
    
    
\end{itemize}

The main findings we obtained are: 

\begin{itemize} 
    \item In general, model-based optimisers (e.g., SMAC~\cite{hutter2011sequential}) outperform model-free optimisers under tight budgets, while model-free optimisers (e.g., GA and IRACE~\cite{lopez2016irace}) become competitive when larger budgets are available. 
    
    \item That said, there exists one optimiser, FLASH~\cite{nair2018finding}, that consistently performs very well on most systems regardless of the budgets. 
    
    

    
    \item The tuning quality is not strongly determined by explicit system attributes, e.g., the search space size and number of options. Instead, other factors, such as the distribution of configurations (in particular, the distribution of local optima), play a crucial role.

\end{itemize}

The rest of this paper is structured as follows. Section~\ref{sec:background} introduces the background and the considered optimisers. Section~\ref{sec:exp_design} details the experimental design. Section~\ref{sec:Results} reports the experimental results and analysis. 
Section~\ref{sec:Implications} provides actionable insights for industry practitioners and configuration tuning researchers. 
Section~\ref{sec:threats} presents the threats to validity and Section~\ref{sec:related_work} reviews related work. Section~\ref{sec:Conclusion} concludes the paper.

\section{Background}\label{sec:background}
\subsection{Software Configuration Tuning Problem}

A configurable software system includes a set of key configuration options, with the $i$th option represented as $x_i$, which can be either a binary or numeric variable~\cite{chen2024mmo}. 
The search space, $\mathcal{X}$, is defined as the Cartesian product of all possible values of $x_i$. 
The aim of tuning the software configuration is to determine optimal performance (e.g., latency, throughput, or accuracy). 
Without loss of generality, the configuration tuning problem can be defined as: 
$\arg\min_{\mathbf{x} \in \mathcal{X}} f(\mathbf{x}),$
where $\mathbf{x} = (x_1, x_2, \ldots, x_n)$ denotes variables (options) in decision space, $n$ denotes the total number of options, and $f$ is the objective function to be optimised. 


\subsection{Optimisers to be Considered}\label{sec:method}

Over the past decades, various optimisers and systems have been proposed to solve configuration tuning problems~\cite{chen2025accuracy}. 
In this study, to make the findings broadly applicable, we investigate the performance of general-purpose optimisers across diverse configurable systems.
We select optimisers based on the four specific criteria: (1) Standalone optimiser, (2) Discrete, (3) General applicability, and (4) Dataset constraints. 
If an optimiser meets all the criteria, it will be pre-selected, followed by further filtering based on its similarity and connections. 
The details of each criterion are as follows:

\begin{itemize}

    
    


    \item \textbf{Optimiser:} This criterion focuses on whether the optimiser is standalone, rather than part of a larger tuning framework (system). In this work, we focus on optimisers themselves, rather than full-fledged tuning systems that integrate multiple techniques~\cite{duan2009tuning,fekry2020tune,zhang2021restune,iqbal2023cameo,cao2024etune,giannakouris2024demonstrating,lao2025gptuner,dong2024mletune,huang2024llmtune,fan2024latuner,li2024large,lee2024k2vtune,yang2024vdtuner,zhan2024knobtune}, e.g., iTuned~\cite{duan2009tuning}, CAMEO~\cite{iqbal2023cameo}, and LATuner~\cite{fan2024latuner}.

    
    \item \textbf{Discrete:} \re{This criterion considers whether an optimiser or system is designed to handle discrete or mixed (i.e., discrete and continuous) configuration spaces. Some otherwise optimisers mainly designed for purely continuous optimisation~\cite{jiang2025trading,jiang2026do}, e.g., vanilla BO (VBO)~\cite{kushner1964new} and CMA-ES~\cite{hansen2001completely}, are excluded.} 

    

    \item \textbf{General applicability:} 
    This criterion focuses on whether an optimiser or system is designed for general-purpose configuration tuning, rather than for a specific type of system or objective (e.g., multi-fidelity optimisers~\cite{li2018hyperband,falkner2018bohb,awad2021dehb}). 
    For example, 
    Optimisers tailored to particular domains, e.g., BOCA for compiler autotuning~\cite{chen2021efficient} and LATuner for database systems~\cite{fan2024latuner}, are excluded to ensure the results are applicable across diverse configurable systems. 

    \item \textbf{\re{Dataset constraints:}} \re{This criterion focuses on whether an optimiser or system is compatible with the datasets considered in this study.}
    \re{It is worth noting that in this study we do not include multi-fidelity methods~\cite{li2018hyperband,falkner2018bohb,awad2021dehb,lindauer2022smac3,pfisterer2022yahpo}, as our datasets contain only single-fidelity evaluations without wall-time or intermediate fidelity information. 
    }
    \re{In addition, we filter out Pareto-optimality-based methods~\cite{lustosa2024learning,chen2023weights,senthilkumar2024can,chen2024adapting} as most systems in the datasets considered in this study contain only a single performance metric. 
    }
    It is worth noting that we do not filter out SWAY, which can be adapted by changing its comparison function without having to use an additional objective~\cite{chen2025accuracy}.
    \re{In addition, we do not consider SPL optimisers~\cite{oh2023finding,hierons2016sip,oh2017finding,guegain2023configuration}, as they require a reliable feature model of the software system, which is not available for most of the systems considered in our study.}

\end{itemize}



Table~\ref{tbl:optimisers} shows the candidate optimisers/systems with their associated venues and the selection criteria. 
After applying these criteria, we initially pre-select the 11 optimisers: Random Search (RS)~\cite{rastrigin1963convergence}, Hill Climbing (HC)~\cite{newell1972human}, Simulated Annealing (SA)~\cite{kirkpatrick1983optimization}, Tabu Search (TS)~\cite{glover1998tabu}, ParamILS~\cite{hutter2009paramils}, GA~\cite{holland1992adaptation}, IRACE~\cite{lopez2016irace}, SWAY~\cite{chen2018sampling}, SMAC~\cite{lindauer2022smac3}, TPE~\cite{bergstra2011algorithms}, and FLASH~\cite{nair2018finding}.
We then further exclude HC, TS, and ParamILS for the following reasons. 
Among the three local search heuristics (HC, SA, and TS), we filter out HC and TS, as SA has been shown to be more effective than both~\cite{nie2011survey,koc2021satune} and competitive with other optimisers~\cite{chen2021multi,koc2021satune}. 
Additionally, we filter out ParamILS as SMAC inherits the main idea of ParamILS and outperforms ParamILS~\cite{chen2024adapting}. 
In fact, the authors of ParamILS and SMAC recommended using SMAC~\cite{pushak2020golden}.


Overall, we consider 8 optimisers, i.e., SMAC, TPE, FLASH, IRACE, RS, SWAY, SA, and GA. Some of them are model-based and the others are model-free.   

\renewcommand{\arraystretch}{0.95} 
\begin{table*}[t]\scriptsize
    \centering
    \rowcolors{2}{blue!10}{white} 
    \caption{Illustration of the representative candidate optimisers/systems with their associated venues and the four selection criteria. 
    Note that if a software configuration optimiser or system meets all the criteria, it is pre-selected. 
    Among the pre-selected optimisers, we exclude HC, TS, and ParamILS: HC and TS have been shown to be less effective than SA~\cite{nie2011survey,koc2021satune}, and the authors of ParamILS and SMAC recommend using SMAC instead~\cite{pushak2020golden}. 
    Finally, the 8 optimisers are selected for investigation in this study: SMAC, TPE, FLASH, IRACE, RS, SWAY, SA, and GA.}
    \resizebox{\textwidth}{!}{%
    \begin{tabular}{lcccccccc}
    \toprule

     \bfseries Optimiser/system (Ref.) & \bfseries Used by & \bfseries Optimiser & \bfseries Discrete &  \bfseries General applicability &  \bfseries Dataset constraints & \bfseries Pre-selected & \bfseries Category & \bfseries Selected \\
     
     \midrule

     \textbf{VBO}~\cite{kushner1964new} 
     & TCC~\cite{guindani2024integrating} & \cmark &        & \cmark &        &  & Model-based\\

     \textbf{SMAC}~\cite{hutter2011sequential,lindauer2022smac3} 
     & PACMSE~\cite{chen2024adapting} & \cmark & \cmark & \cmark & \cmark & \cmark & Model-based & \cmark \\
     
     \textbf{TPE}~\cite{bergstra2011algorithms} 
     & TOSEM~\cite{lustosa2024learning} & \cmark & \cmark & \cmark & \cmark & \cmark & Model-based & \cmark \\
     
     \textbf{FLASH}~\cite{nair2018finding}
     & TSE~\cite{nair2018finding} & \cmark & \cmark & \cmark & \cmark & \cmark & Model-based & \cmark \\

     \textbf{BOCA}~\cite{chen2021efficient}
     & ICSE~\cite{chen2021efficient} & \cmark & \cmark &       & &  & Model-based \\

    
     \textbf{CAMEO}~\cite{iqbal2023cameo}
     & SoCC~\cite{iqbal2023cameo} &      & \cmark & \cmark  & \cmark &  & Model-based & \\

     \textbf{IRACE}~\cite{lopez2016irace}
     & ICSE~\cite{chen2021efficient} & \cmark & \cmark & \cmark & \cmark 
     & \cmark & Model-free (racing-based)
     & \cmark \\
     
     \textbf{RS}~\cite{rastrigin1963convergence} 
     & TOSEM~\cite{lustosa2024learning} & \cmark & \cmark & \cmark & \cmark & \cmark & Model-free (sampling-based) & \cmark \\

     \re{\textbf{RRS}~\cite{oh2023finding}}
     & \re{TOSEM~\cite{oh2023finding}} & \cmark & \cmark &  &  & & \re{Model-free (sampling-based)} &  \\
     
     \textbf{SWAY}~\cite{chen2018sampling}
     & TSE~\cite{chen2018sampling} & \cmark & \cmark & \cmark & \cmark & \cmark & Model-free (sampling-based) & \cmark \\

     \textbf{HC}~\cite{newell1972human}
     & TSE~\cite{chen2024mmo} & \cmark & \cmark & \cmark & \cmark & \cmark & Model-free (local search heuristic) & \\
     
     \textbf{SA}~\cite{kirkpatrick1983optimization}
     & TOSEM~\cite{chen2023weights} & \cmark & \cmark & \cmark & \cmark & \cmark & Model-free (local search heuristic) & \cmark \\
     
     \textbf{TS}~\cite{glover1998tabu} 
     & ASE~\cite{diaz2003automated} & \cmark & \cmark & \cmark & \cmark & \cmark & Model-free (local search heuristic) & \\

     
     \textbf{ParamILS}~\cite{hutter2009paramils} 
     & PACMSE~\cite{chen2024adapting} & \cmark & \cmark & \cmark & \cmark & \cmark & Model-free (local search heuristic) & \\

     \textbf{GA}~\cite{holland1992adaptation}
     & FSE~\cite{chen2021multi} & \cmark & \cmark & \cmark & \cmark & \cmark & Model-free (global search heuristic) & \cmark \\

    \bottomrule
     
    \end{tabular}
    }
    \label{tbl:optimisers}
\end{table*}

\subsubsection{Model-based optimisers}

To guide the tuning, model-based optimisers do not rely solely on the system measurements, but also on a gradually updated surrogate model that can cheaply predict the performance. 
Typically, they follow the main idea and procedure of BO: a surrogate model is trained on observed data to predict the performance of configurations; then an acquisition function is optimised to select the next configuration to evaluate. In this work, we investigate the three model-based optimisers, namely: 

\begin{itemize}
    \item \textbf{SMAC~\cite{hutter2011sequential}}: SMAC uses a Random Forest model as the surrogate model and employs Expected Improvement as the acquisition function~\cite{lindauer2022smac3}. 

    \item \textbf{FLASH~\cite{nair2018finding}}: FLASH uses the Classification and Regression Tree (CART)~\cite{breiman1984classification} as the surrogate model and employs the Maximum Mean as the acquisition function.

    \item \textbf{TPE~\cite{bergstra2011algorithms}}: TPE models the relationship between configurations and their performance using two non-parametric density estimators, and employs Expected Improvement as the acquisition function. 
    
        
    
    
\end{itemize}

\subsubsection{Model-free optimisers}

Model-free optimisers tune the configuration without building an explicit model~\cite{chen2024adapting}. 
In this work, we investigate five model-free optimisers:

\begin{itemize}
    \item \textbf{RS~\cite{rastrigin1963convergence}}: RS randomly generates configurations and measures their performance, which is used as a baseline. 

    \item \textbf{SA~\cite{kirkpatrick1983optimization}}: Inspired by the annealing process in metallurgy, SA explores the search space by probabilistically accepting worse configurations to escape local optima. 
    
    
    \item \textbf{GA~\cite{holland1992adaptation}}: GA is a population-based optimiser that evolves configurations through the processes of selection, crossover, and mutation.

    \item \textbf{IRACE~\cite{lopez2016irace}}: IRACE is designed to automatically set algorithmic parameters based on an iteratively updated probabilistic model of the search space, refining the distribution of promising options through racing mechanisms. 
    


    \item \textbf{SWAY~\cite{chen2018sampling}}: SWAY is a sampling-based optimiser that generates a very large set of candidate configurations, then uses recursive binary partitioning and representative evaluation to gradually filter out poorly performing configurations, retaining only the best subset.
        
\end{itemize}

\begin{figure}[t]
    \centering
    \includegraphics[width=0.9\linewidth]{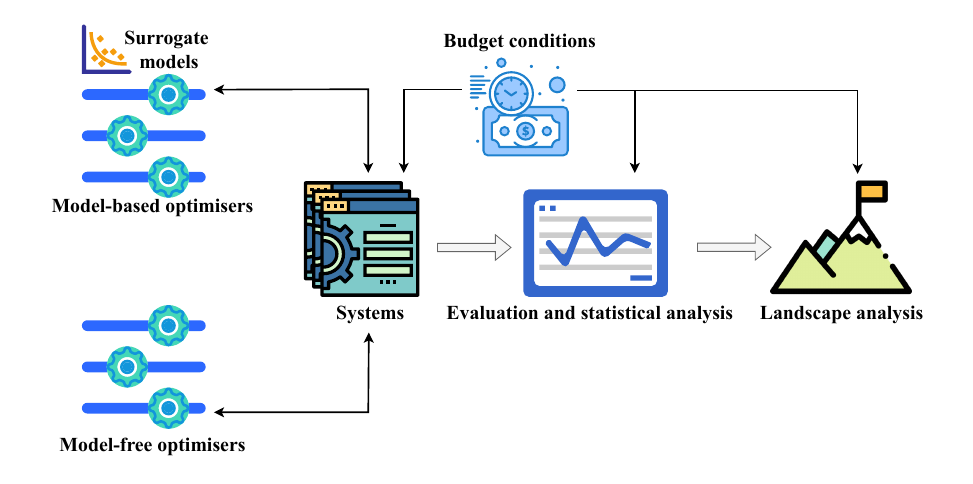}
    
    \captionsetup{skip=10pt}
    \caption{The workflow of our empirical study. 
    }
    \label{fig:workflow}
\end{figure}

\section{Experimental Design}\label{sec:exp_design}

The workflow of our work is given in Figure~\ref{fig:workflow}. 
We evaluate the eight model-based and model-free optimisers for tuning 22 configurable systems, in which model-based optimisers utilise surrogate models to guide the tuning. 
Each tuning experiment is conducted over 30 repeated runs. 
The results obtained under different budget conditions are then evaluated and analysed using the evaluation metrics and statistical validation methods outlined in Sections~\ref{Sec:EM} and~\ref{Sec:SV}. 
In addition, we analyse and discuss why different optimisers perform differently under varying budget conditions through fitness landscape analysis, with the employed metrics detailed in Section~\ref{sec:RQ2_method}.

\subsection{Research Questions}

Our experimental evaluation seeks to answer three research questions (\textbf{RQs}). 
First, we would like to investigate how each optimiser performs under different budget levels within a commonly used evaluation range. To explore this, we seek to answer:

\begin{tcolorbox}[boxrule=1pt, left=2pt, right=2pt, top=2pt, bottom=2pt]  
    \textbf{RQ1:} \textit{How does tuning quality vary under commonly used budget ranges?} 
\end{tcolorbox}

In some cases, software practitioners may be allowed to have more flexible budgets (up to ten thousand evaluations~\cite{acher2019learning,oh2023finding}).
Hence, we seek to answer:

\begin{tcolorbox}[boxrule=1pt, left=2pt, right=2pt, top=2pt, bottom=2pt] 
\textbf{RQ2:} \textit{What happens when larger budgets are available?}
\end{tcolorbox}

While \textbf{RQ1} and \textbf{RQ2} can reveal what happens under different budget conditions, they do not explain why such differences arise. This naturally leads to our third \textbf{RQ}:

\begin{tcolorbox}[boxrule=1pt, left=2pt, right=2pt, top=2pt, bottom=2pt] 
\textbf{RQ3:} \textit{How can the answers to \textbf{RQ1} and \textbf{RQ2} be explained?} 
\end{tcolorbox}

\begin{table*}[!ht]\scriptsize
    \centering
    \rowcolors{2}{blue!10}{white} 
    \caption{Subject configurable software systems (ordered by the number of configurations), along with their performance objectives studied. In the table, the symbol $O$ denotes the number of options (dimensionality) and $S$ denotes the configuration space.}
    
    \resizebox{\textwidth}{!}{%
    \begin{tabular}{ccccccccc}
    \hline
    \textbf{System} & \textbf{Version} & \textbf{Workload} &
    \textbf{Language} & \textbf{Domain} & \textbf{Performance} &
    \textbf{$O$} & \textbf{$S$} & Used by \\
    \hline

    \textsc{Brotli} & 1.0.7 & Compressing a 1 GB file & C & Compression & Runtime & 2 & $1.80 \times 10^2$ & \cite{weber2023twins,chen2025accuracy} \\
    \textsc{Apache} & 21.0 & ApacheBench 2.3 & C & Web Server & Runtime & 6 & $3.20 \times 10^2$ & \cite{weber2023twins,chen2025accuracy} \\
    \textsc{HSQLDB} & 19.0 & PolePosition 0.6.0 & Java & Database & Runtime & 7 & $8.64 \times 10^2$ & \cite{weber2023twins,chen2025accuracy} \\
    \textsc{PostgreSQL} & 22.0 & PolePosition 0.6.0 & C & Database & Runtime & 8 & $8.64 \times 10^2$ & \cite{weber2023twins,chen2025accuracy} \\
    \textsc{Storm} & 0.9.5 & Randomly generated benchmark & Clojure & Streaming Process & Latency & 12 & $4.09 \times 10^3$ & \cite{krishna2020whence} \\
    \textsc{Z3} & 4.8.14 & AUFNIRA & C/C++ & SMT Solver & Runtime & 12 & $4.09 \times 10^3$ & \cite{muhlbauer2023analysing,ye2025distilled} \\
    \textsc{x264} & 0.152.2851 & Encoding `Sintel' 480p (y4m) to H.264 & C & Video Encoder & Runtime & 10 & $4.61 \times 10^3$ & \cite{weber2023twins} \\
    \textsc{LRZIP} & 0.631 & Compressing a 621 MB file & C++ & File Compressor & Runtime & 7 & $5.18 \times 10^3$ & \cite{weber2023twins} \\
    \textsc{MongoDB} & 4.0.1 & PolePosition 0.6.0 & C++ & Database & Runtime & 8 & $6.84 \times 10^3$ & \cite{weber2023twins,chen2025accuracy} \\
    \textsc{BATIK} & 1.14 & Corona & Java & SVG Rasterizer & Runtime & 11 & $1.22 \times 10^4$ & \cite{muhlbauer2023analysing,ye2025distilled} \\
    \textsc{NGINX} & 1.14.0 & ApacheBench 2.3 & C & Web Server & Runtime & 11 & $1.53 \times 10^4$ & \cite{weber2023twins,gong2024predicting} \\
    \textsc{Spear} & 2.7 & Randomly generated benchmark & C & SAT Solver & Runtime & 14 & $1.64 \times 10^4$ & \cite{krishna2020whence} \\
    \textsc{SQLite} & 3.19.0.0 & Randomly generated benchmark & C & Database & Runtime & 14 & $1.63 \times 10^4$ & \cite{krishna2020whence} \\
    \textsc{H2} & 1.4.200 & Smallbank-1 & Java & Database & Throughput & 16 & $6.55 \times 10^4$ & \cite{muhlbauer2023analysing,ye2025distilled} \\
    \textsc{LLVM} & 6.0.0 & All cpp files of TMV & C & Optimiser & Runtime & 16 & $6.55 \times 10^4$ & \cite{weber2023twins} \\
    \textsc{7z} & 9.20 & Compressing a 3 GB directory & C++ & Compression & Runtime & 8 & $6.86 \times 10^4$ & \cite{weber2023twins,chen2025accuracy} \\
    \textsc{ExaStencils} & 1.2 & Three default benchmarks & Scala & Code Generator & Runtime & 7 & $1.07 \times 10^5$ & \cite{weber2023twins,gong2024predicting} \\
    \textsc{DCONVERT} & 1.0.0-$\alpha$7 & Jpeg large & Java & Image Scaling & Runtime & 18 & $2.62 \times 10^5$ & \cite{muhlbauer2023analysing,ye2025distilled} \\
    \textsc{Jump3r} & 1.0.4 & Beethoven & Java & Audio Encoder & Runtime & 16 & $2.94 \times 10^5$ & \cite{muhlbauer2023analysing,ye2025distilled} \\
    \textsc{Kanzi} & 5.0 & All cpp files of TMV & Java & Compression & Runtime & 18 & $2.35 \times 10^6$ & \cite{weber2023twins,chen2025accuracy} \\
    \textsc{XZ} & 5.2.0 & Ambivert & C/C++ & File Compressor & Runtime & 33 & $8.58 \times 10^9$ & \cite{muhlbauer2023analysing,ye2025distilled} \\

    \re{\textsc{VP9}} & \re{1.0} & \re{Encoding videos} & \re{C} &
    \re{Video Encoder} & \re{Runtime} & \re{42} &
    \re{$8.59 \times 10^9$} & \re{\cite{oh2023finding}} \\
    
    \hline
    \end{tabular}
    }

    \label{tbl:systems}
\end{table*}

\subsection{Configurable Software Systems}\label{sec:configsystems}

We select software systems based on the following criteria: (1) Systems have been studied in key venues in software and system engineering; (2) When multiple sets of relevant configuration options are available for a system, we prioritise the one with the highest number of options, following the practice~\cite{chen2025accuracy}. For instance, \textsc{Storm} has been extensively studied in prior research, and we select the instance with 12 options, representing the most complex case. 
Following these criteria, we consider the 22 highly configurable software systems that have been widely used in existing work~\cite{muhlbauer2023analysing,krishna2020whence,weber2023twins,cao2023cm,ye2025distilled,chen2025accuracy}, including the same sets of performance objectives, options, and possible configurations, as shown in Table~\ref{tbl:systems}. 
For the 9 systems, i.e., \textsc{Brotli}, \textsc{Apache}, \textsc{HSQLDB}, \textsc{PostgreSQL}, \textsc{x264}, \textsc{LRZIP}, \textsc{Spear}, \textsc{LLVM}, and \textsc{7z}, all possible configurations are present in the dataset. 
For the remaining systems, if a configuration to be measured does not exist in the dataset, its performance is estimated based on the nearest neighbour’s performance, following the practice~\cite{chen2025accuracy,ye2025distilled,ganguly2025bingo}.
Note that the dataset we considered was constructed based on repeated measurements for each configuration, obtained from~\cite{muhlbauer2023analysing,krishna2020whence,weber2023twins}, and the median/mean was used as the final performance value.





\subsection{Optimiser Settings}

We implement SA, GA, IRACE, SMAC, and TPE through the Python libraries simanneal\footnote{\url{https://github.com/perrygeo/simanneal}}, DEAP\footnote{\url{https://github.com/DEAP/deap}}, irace\footnote{\url{https://github.com/MLopez-Ibanez/irace}}, SMAC3\footnote{\url{https://github.com/automl/SMAC3}}, and hyperopt\footnote{\url{https://github.com/hyperopt/hyperopt}}, respectively. 
Furthermore, the implementations of FLASH and SWAY are adapted from the authors' GitHub repositories\footnote{\url{https://github.com/FlashRepo} and \url{https://github.com/ginfung/FSSE}}.

For SA, the initial temperature is set to 10, following the practice in~\cite{chen2023weights}. 
We consider the settings of GA that were widely used in prior works~\cite{trotter2019forecasting,chen2021multi,chen2024mmo}.
Specifically, we employ the binary tournament for mating selection and a population size of 50, together with the boundary mutation and uniform crossover under the rates of 0.1 and 0.9, respectively. 
For model-based optimisers, i.e., SMAC, TPE and FLASH, we set the initial sampling budget of 30 according to~\cite{nair2018finding}. 
Additionally, for FLASH, we allow 1,000 surrogate evaluations for predictions, as recommended in~\cite{chen2024mmo,chen2025accuracy}.

\subsection{Tuning Budget}

Apart from SWAY, the maximum search budget for all the optimisers is set to $\min (10^4, S)$, denoted as $N$, where $S$ is the size of the search space (i.e., all possible configurations). 
For SWAY, following the authors' setting~\cite{chen2018sampling}, we sample $N$ initial configurations (without measurements) and set the total search budget to $\sqrt{N}$. 
Note that SWAY samples $N$ candidate configurations (e.g.,10,000) and recursively prunes them by measuring two configurations per split (similar to binary search), resulting in a small evaluation budget. 
It is worth noting that some optimisers may get trapped in local optima, preventing them from discovering new configurations to fully utilise the allocated search budget. 
To mitigate this issue, we empirically set a termination condition of 2 hours as we found that even with additional time, e.g. more than 2 hours, the performance improvement remains negligible. 
Note that most optimisers reach the maximum search budget (e.g.,10,000 distinct configurations) within half an hour, while some optimisers (e.g.,SMAC) may stop producing new configurations after about one hour on some systems. 
The search budget of 2 hours is also recommended in~\cite{chen2021multi,jamshidi2016uncertainty}. 
To allow statistical comparison, each tuning was repeated 30 times. 
In each tuning run, we ensure that only distinct configurations count towards the budget consumption, following the common practice~\cite{jamshidi2016uncertainty,nair2018finding,chen2021multi,chen2025accuracy}.

\subsection{Evaluation Metrics}\label{Sec:EM}

In this work, we directly use performance values as a metric, along with their means and standard deviations. 
Additionally, since we know the optimum of each tuning problem, we also adopt the well-known metric, i.e., regret, which quantifies the difference between the best performance value obtained by an optimiser and the true optimum value. 

\subsection{Statistical Validation}\label{Sec:SV}

We use the Wilcoxon rank-sum test~\cite{wilcoxon1992individual} with the significance level of $\alpha=0.05$ and Holm-Bonferroni correction~\cite{holm1979simple} to determine if two optimisers are statistically different over 30 repeated runs. 
Note that if an optimiser is not statistically worse than any other optimiser, it is considered statistically the best, so there can be multiple best optimisers for a setup.

\begin{figure}[!ht]
    \centering
    \begin{minipage}{0.7\linewidth}
        \centering
        \includegraphics[width=0.8\linewidth]{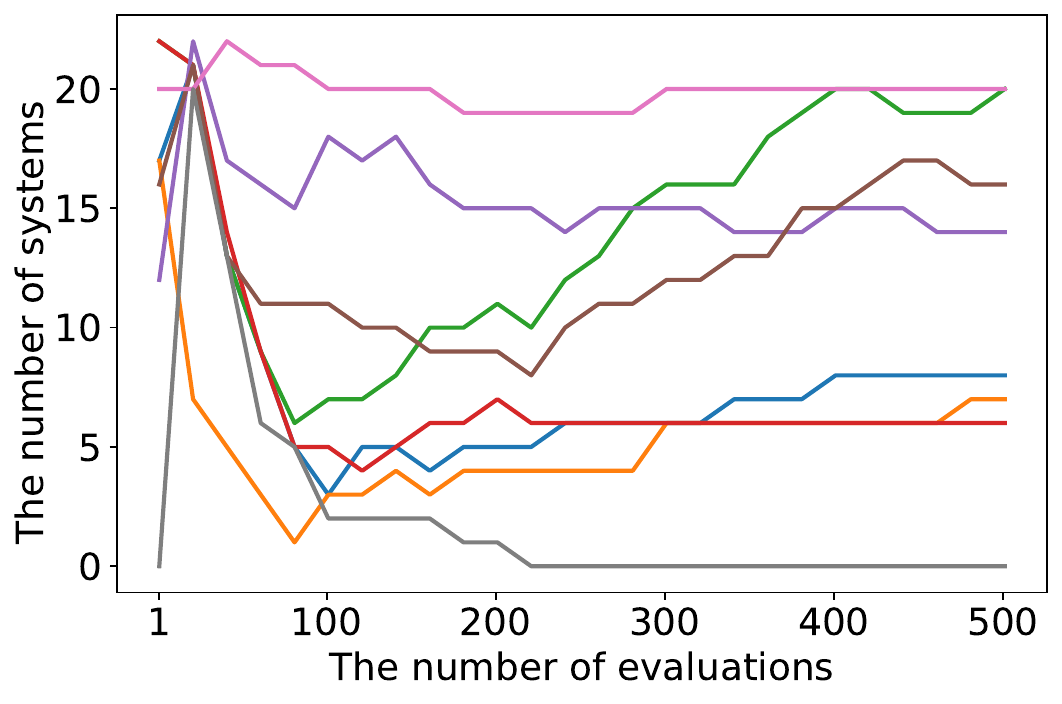}
    \end{minipage}
    \begin{minipage}{0.8\linewidth}
        \centering
        \includegraphics[width=0.95\linewidth]{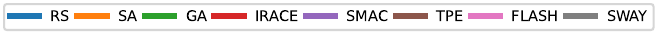}
    \end{minipage}

    \captionsetup{skip=4pt}
    \caption{Trajectories of the number of systems where each optimiser is statistically the best out of the 22 systems over 30 independent runs, with respect to the varying budget of up to 500 evaluations.
    }
    \label{fig:percentage500}
\end{figure}

\vspace{-5pt}
\section{Results and Analysis}\label{sec:Results}

We now present and discuss the experiment results.

\subsection{How Does Tuning Quality Vary under Commonly Used Budget Ranges?}\label{sec:RQ1}

\subsubsection{Method}


To answer \textbf{RQ1}, we track the performance of the 8 optimisers across commonly used budget ranges (up to 500 evaluations) on the 22 systems. 
The reason we set 500 evaluations is based on prior studies, most of which have a budget below or equal to it~\cite{ye2025distilled,chen2021multi,chen2024adapting,jamshidi2016uncertainty,nair2018finding,chen2025accuracy,chen2021efficient,chen2018sampling,chen2024mmo,lustosa2024learning,huang2025rethinking,iqbal2023cameo,huang2024llmtune,li2024large,senthilkumar2024can}.


\subsubsection{Findings}

Figure~\ref{fig:percentage500} shows trajectories of the number of systems where each optimiser is statistically the best out of the 22 systems over 30 independent runs, with respect to the varying evaluation budget of up to 500 evaluations.

As shown in Figure~\ref{fig:percentage500}, the model-based optimiser FLASH consistently performs best on at least 19 out of 22 systems, regardless of the evaluation budget within this range. 
Among the other model-based optimisers, SMAC performs well, being the best optimiser on at least 14 systems (peaking at 18) across the varying evaluation budget, while the performance of TPE gradually improves as the budget increases (peaking at 18). 
Although neither SMAC nor TPE surpasses FLASH, both outperform all the model-free optimisers when the budget is highly limited, e.g., at 100 evaluations.


\begin{quotebox}
   \noindent
   \textit{\textbf{Finding 1:} 
   The common belief that model-based optimisers (e.g., FLASH and SMAC) outperform model-free optimisers under limited budgets (e.g., 100 evaluations) generally holds \re{in software configuration tuning}. }
\end{quotebox}

Among the model-free optimisers, GA is the most competitive optimiser compared with RS, SA, IRACE and SWAY. 
While GA initially lags behind the model-based optimisers, it gradually catches up as the budget increases. 
\re{Notably, after 300 evaluations, GA begins to surpass most other optimisers, except FLASH, and becomes one of the best-performing methods. }


\begin{quotebox}
   \noindent
   \textit{\textbf{Finding 2:} Among model-free optimisers, the global search strategy (GA) may be more effective than local search strategies (e.g., SA and IRACE) and sampling-based strategies (e.g., RS and SWAY). 
   Additionally, the global search strategy is beneficial when a relatively large budget is available. 
   }
\end{quotebox}

\begin{figure}[!ht]  
    \centering
    \renewcommand{\thesubfigure}{}  
    
    \begin{minipage}{0.36\linewidth}
        \includegraphics[width=\linewidth]{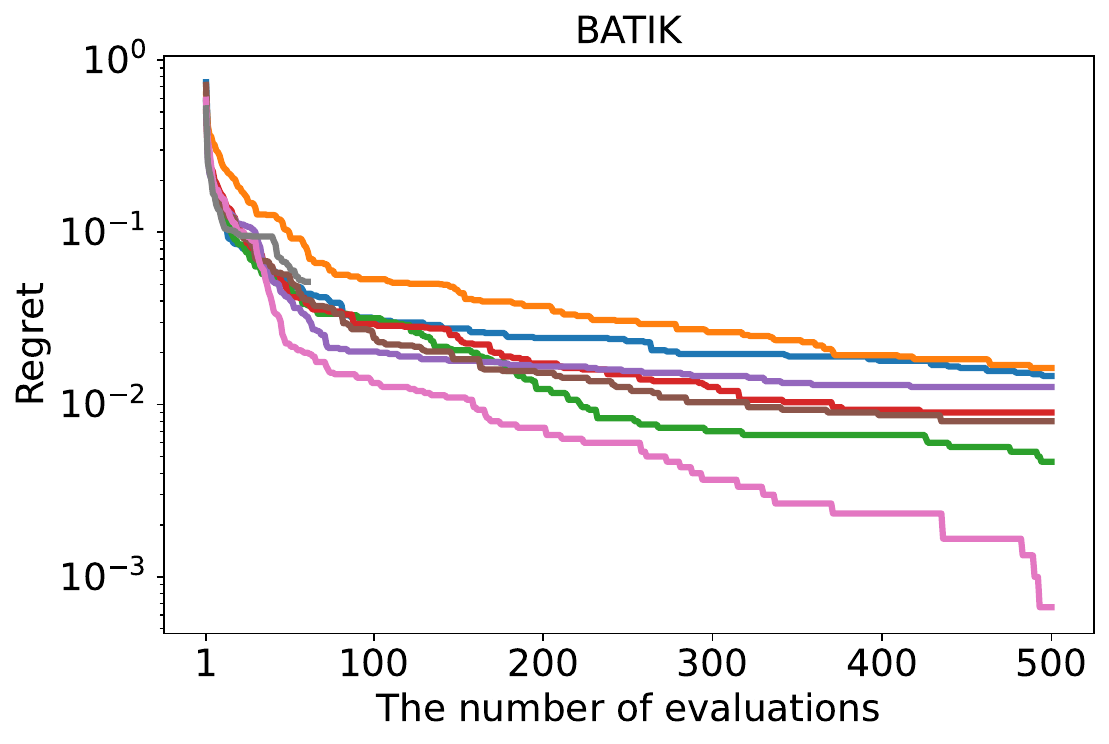}
    \end{minipage}
    \begin{minipage}{0.36\linewidth}
        \includegraphics[width=\linewidth]{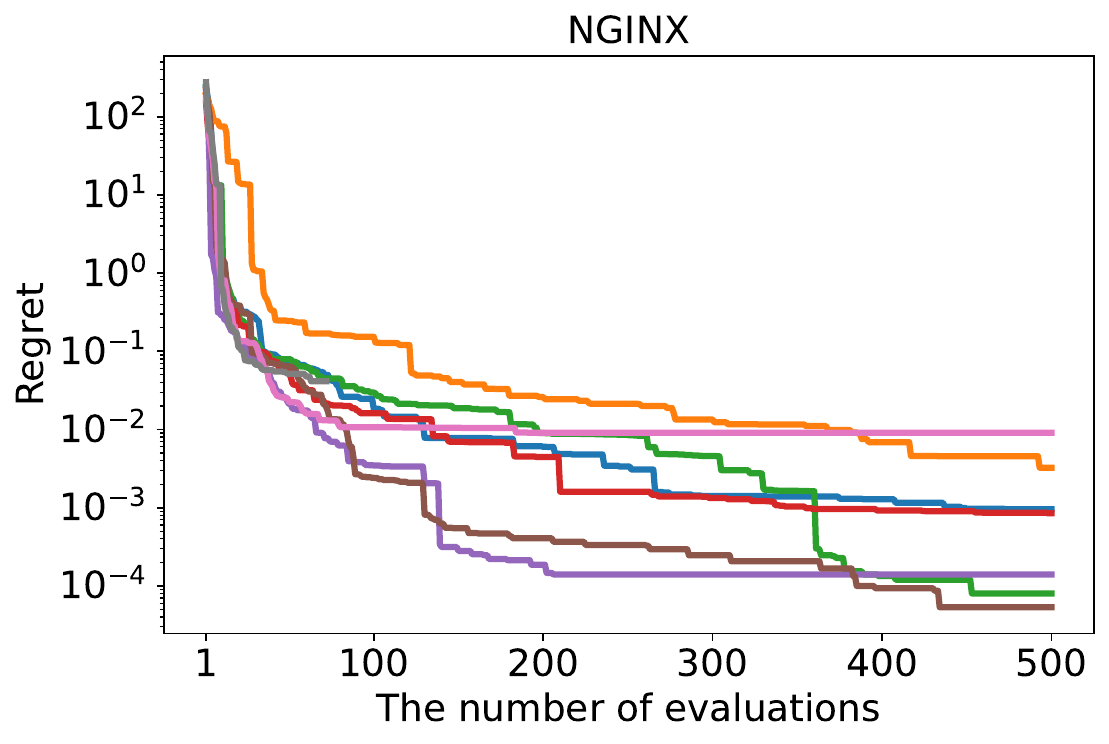}
    \end{minipage}
    \hfill
    \begin{minipage}{0.36\linewidth}
        \includegraphics[width=\linewidth]{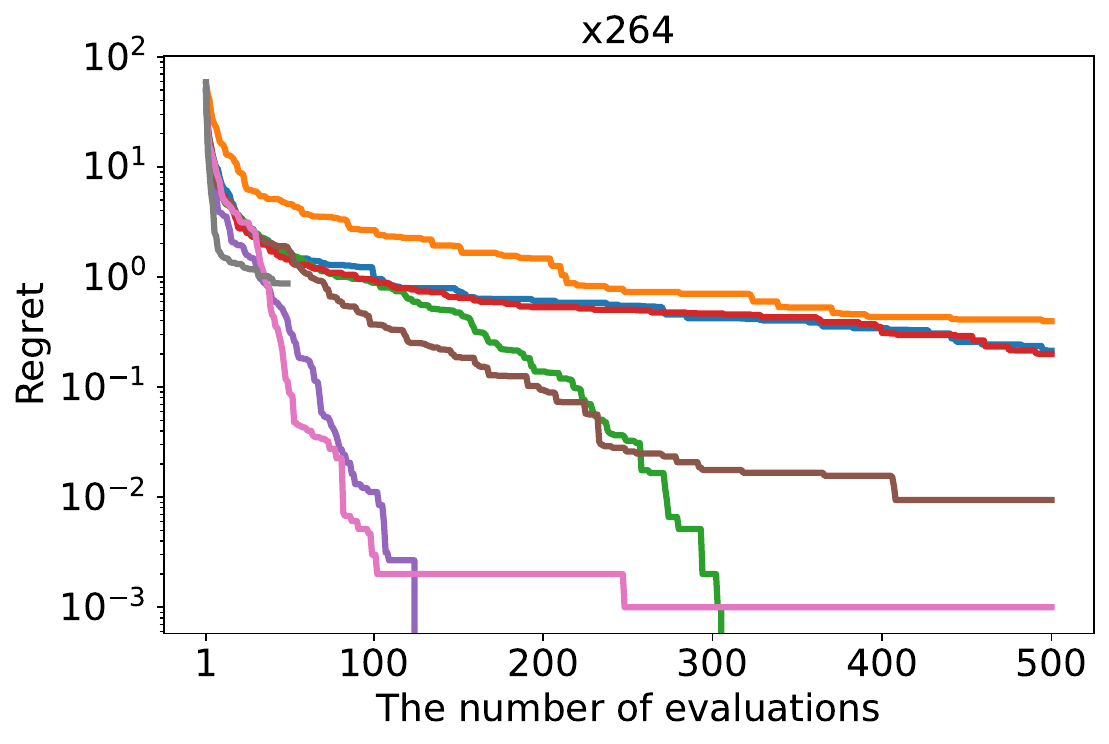}
    \end{minipage}
    \begin{minipage}{0.36\linewidth}
        \includegraphics[width=\linewidth]{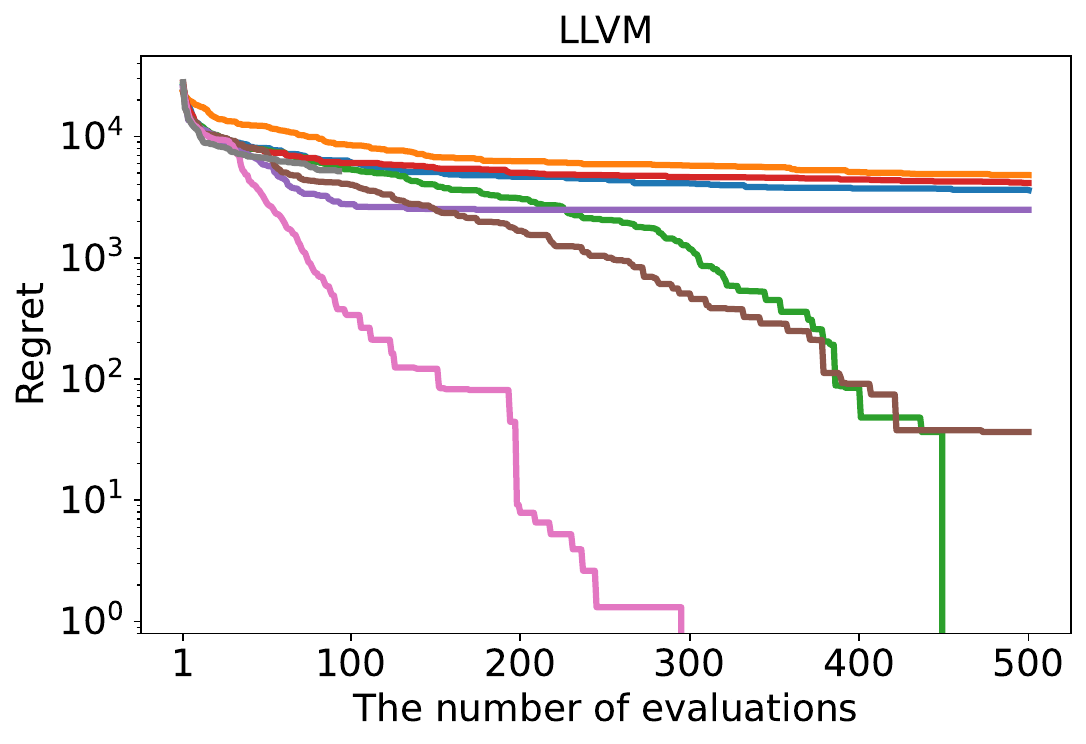}
    \end{minipage}

    
    \begin{minipage}{0.8\linewidth}
        \centering
        \includegraphics[width=0.95\linewidth]{figures/legend.pdf}
    \end{minipage}

    \captionsetup{skip=4pt}
    \caption{Convergence trajectories of the 8 optimisers on the 4 representative systems (\textsc{BATIK}, \textsc{NGINX}, \textsc{x264}, and \textsc{LLVM}) during the tuning process with a varying budget of up to 500 evaluations. Each coloured line represents the mean difference between the best performance obtained and the problem’s true optimum on 30 independent runs.}

    \label{fig:convergence500}
\end{figure}

For local optimisers, i.e., SA and IRACE, both perform best on approximately 6 systems, indicating that they may not be particularly effective for configuration tuning problems under limited budgets. 
Finally, among sampling-based optimisers, RS gradually improves as the budget increases, ultimately achieving the best performance on 8 systems. 
In contrast, SWAY fails to be the best on any system within the budget range. One possible reason for this observation is that RS can explore the search space more evenly as the budget grows, while SWAY’s limited internal budget constrains its ability to thoroughly search the space.

To help understand the search behaviour of the 8 optimisers, we provide their convergence trajectories on the 4 representative systems, i.e., \textsc{BATIK}, \textsc{NGINX}, \textsc{x264}, and \textsc{LLVM}, shown in Figure~\ref{fig:convergence500}. 
The convergence trajectories on the other 18 systems can be found in the supplementary material\footnote{\textcolor{blue}{\url{https://anonymous.4open.science/r/Config-W2W-98B2/Supplementary.pdf}}}. 
For \textsc{BATIK} and \textsc{LLVM}, FLASH shows dominant performance and GA is able to catch up as the evaluation budget increases. 
For \textsc{x264}, GA performs better than FLASH when the evaluation budget is greater than 300. 
In contrast, on \textsc{NGINX}, both FLASH and GA perform worse than TPE.



\begin{figure}[!ht]  
    \centering
    \renewcommand{\thesubfigure}{}  
    
    \begin{minipage}{0.36\linewidth}
        \includegraphics[width=\linewidth]{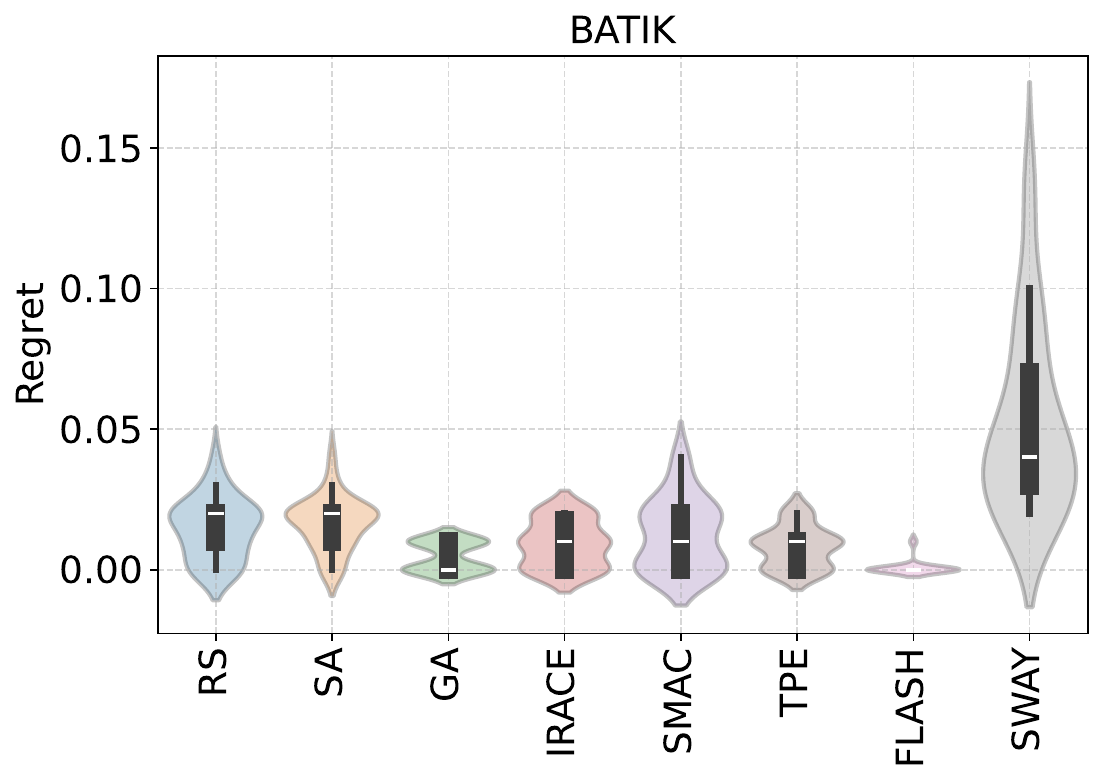}
    \end{minipage}
    \begin{minipage}{0.36\linewidth}
        \includegraphics[width=\linewidth]{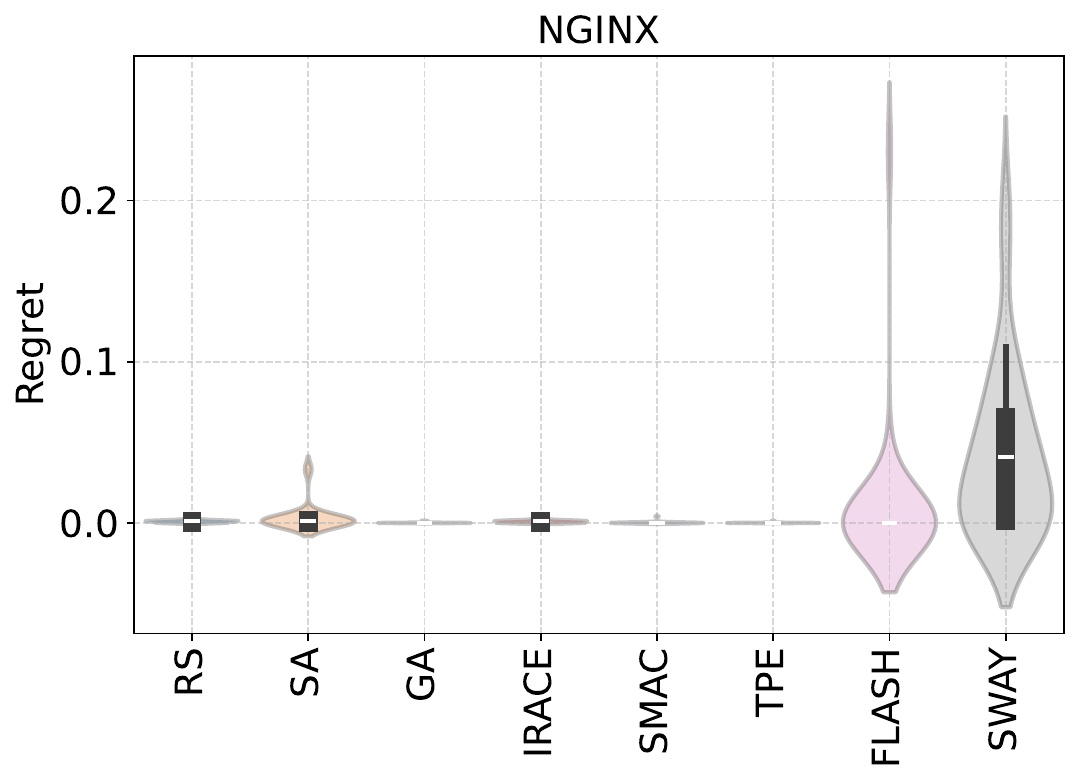}
    \end{minipage}

    \begin{minipage}{0.36\linewidth}
        \includegraphics[width=\linewidth]{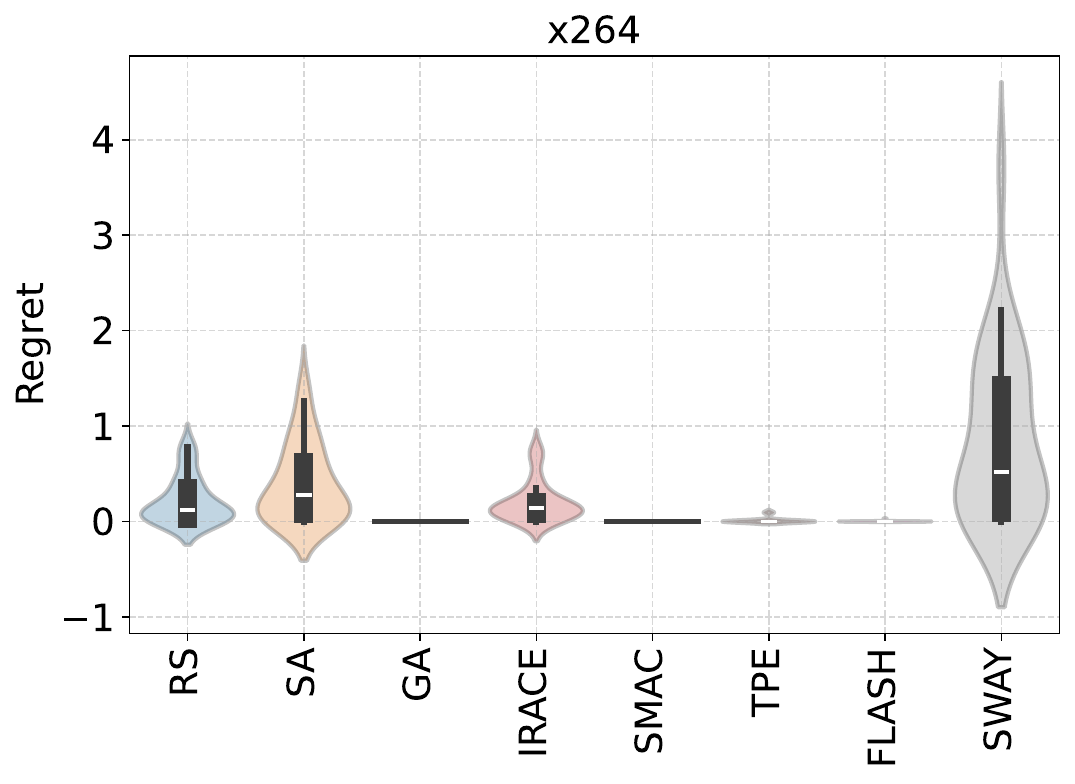}
    \end{minipage}
    \begin{minipage}{0.36\linewidth}
        \includegraphics[width=\linewidth]{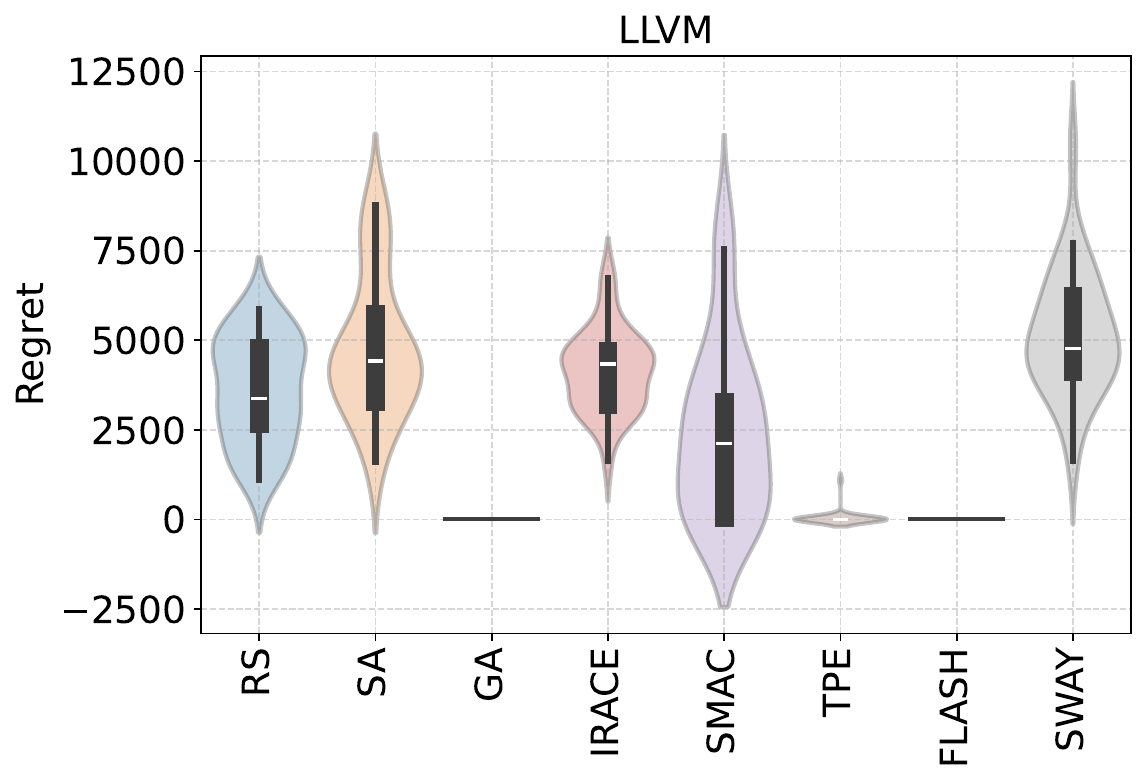}
    \end{minipage}

    \captionsetup{skip=4pt}
    \caption{Distribution of regrets for the 4 representative systems (\textsc{BATIK}, \textsc{NGINX}, \textsc{x264}, and \textsc{LLVM}) across 30 independent runs under the budget of 500 evaluations. The plots highlight the variability in performance among the 8 optimisers, with wider sections indicating higher result density.
    }

    \label{fig:violin_500}
\end{figure}

To further analyse the search behaviour of the 8 optimisers, Figure~\ref{fig:violin_500} shows the distribution of regret values for the 4 systems (\textsc{BATIK}, \textsc{NGINX}, \textsc{x264}, and \textsc{LLVM}) under the budget of 500 evaluations. 
The regret distribution on the other 18 systems can be found in the supplementary material. 
As shown in Figure~\ref{fig:violin_500}, for \textsc{BATIK}, FLASH achieves the lowest median regret (white line in box plot), followed by GA. 
In contrast, FLASH performs poorly on \textsc{NGINX}, which may be due to its greedy strategy. 
For \textsc{x264}, GA and SMAC find the true optimum, while FLASH and TPE come very close. 
In the case of \textsc{LLVM}, both FLASH and GA successfully locate the true optimum, achieving zero regret across 30 runs. 
This suggests that the greedy mechanism of FLASH may not be beneficial for all systems but is generally more effective than optimisers which consider uncertainty, e.g., SMAC.

\begin{quotebox}
   \noindent
   \textit{\textbf{Finding 3:} 
   FLASH performs consistently well on most systems on a range of budget levels, up to 500 evaluations. Additionally, in model-based optimisers, the greedy strategy may be more effective than uncertainty-aware strategies (e.g., SMAC). 
   }
\end{quotebox}


Based on the above analysis, we can conclude that:

\begin{tcbitemize}[raster columns=1, raster rows=1]
    \tcbitem[myhbox={}{Answer to RQ1}]  
    \textit{Under the commonly used tight budget level, model-based algorithms generally perform better than model-free algorithms. In particular, FLASH shows its clear advantage, achieving the best in at least 19 out of the 22 systems.}
\end{tcbitemize}




\subsection{What Happens When Larger Budgets Are Available?}\label{sec:RQ2}

\subsubsection{Method}

To answer \textbf{RQ2}, we track the performance of the 8 optimisers across varying budgets of up to 10,000 evaluations on the 22 systems. 


\subsubsection{Findings}

\begin{figure}[!ht]
    \centering
    \begin{minipage}{0.7\linewidth}
        \centering
        \includegraphics[width=0.8\linewidth]{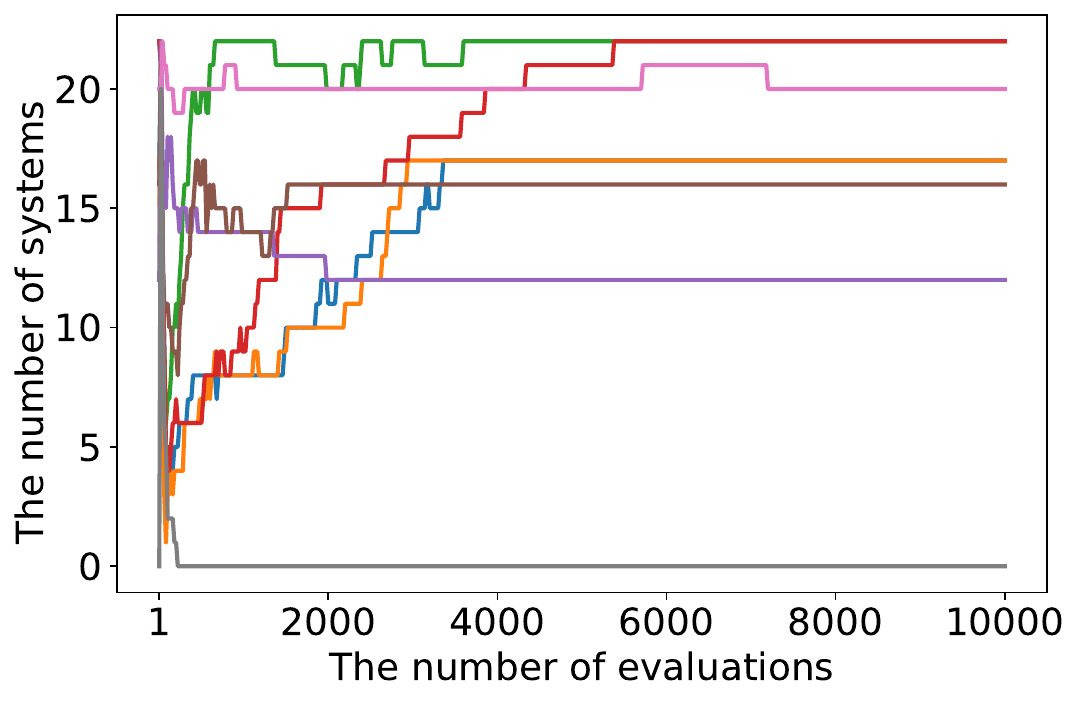}
    \end{minipage}
    \begin{minipage}{0.8\linewidth}
        \centering
        \includegraphics[width=0.95\linewidth]{figures/legend.pdf}
    \end{minipage}

    \captionsetup{skip=4pt}
    \caption{Trajectories of the number of systems where each optimiser is statistically the best out of the 22 systems over 30 independent runs, with respect to the varying budget of up to  10,000 evaluations. 
    }
    \vspace{-10pt}
    \label{fig:percentage10000}
\end{figure}

Figure~\ref{fig:percentage10000} shows trajectories of the number of systems where each optimiser is statistically the best out of the 22 systems over 30 independent runs, with respect to the varying evaluation budget of up to 10,000 evaluations.

As shown in Figure~\ref{fig:percentage10000}, FLASH exhibits stable performance as the evaluation budget increases, performing best in about 20 systems over time. 
In contrast, GA and IRACE consistently improve as the budget increases, reaching the highest number of statistically best results (22 systems) when given generous evaluations, e.g., 5,000.

\begin{figure}[!ht]  
    \centering
    \renewcommand{\thesubfigure}{}  

    \begin{minipage}{0.48\linewidth}
        \includegraphics[width=\linewidth]{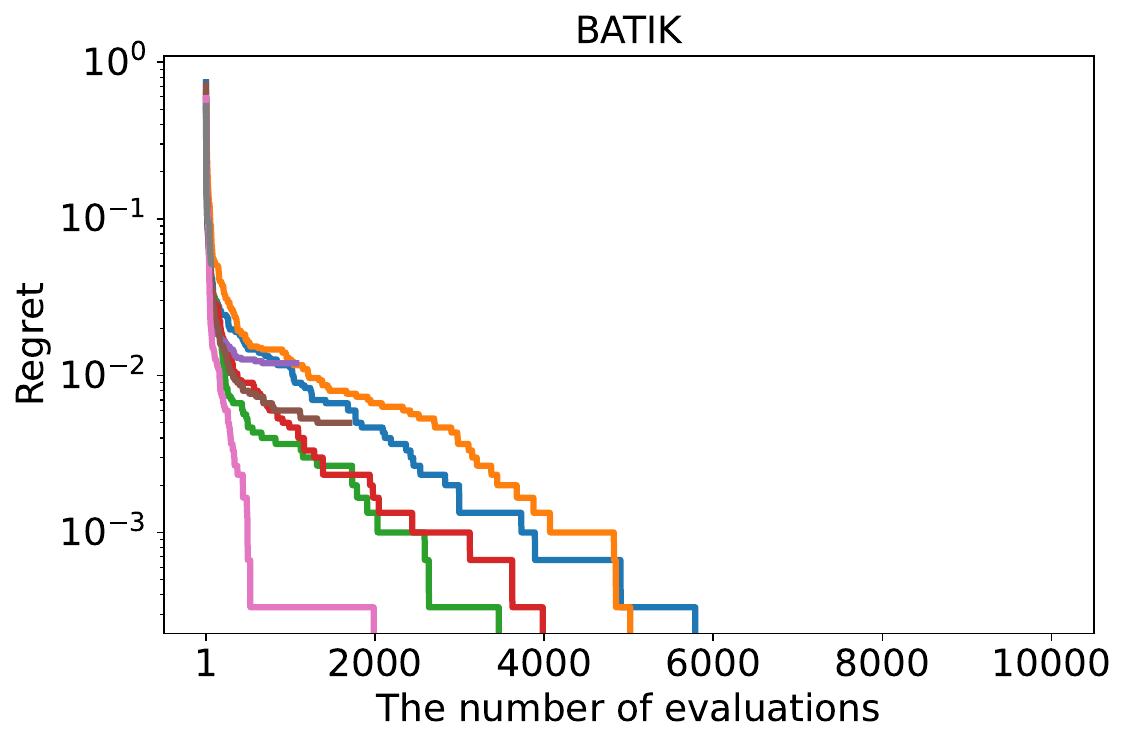}
    \end{minipage}
    \begin{minipage}{0.48\linewidth}
        \includegraphics[width=\linewidth]{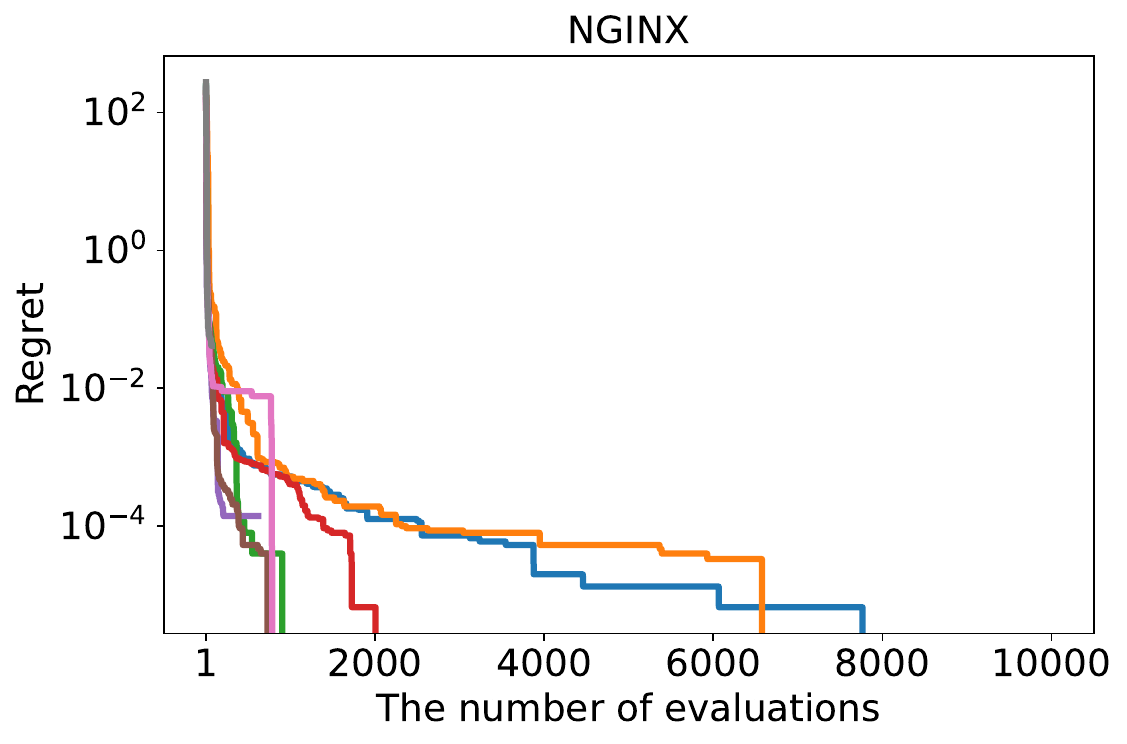}
    \end{minipage}

    \begin{minipage}{0.48\linewidth}
        \includegraphics[width=\linewidth]{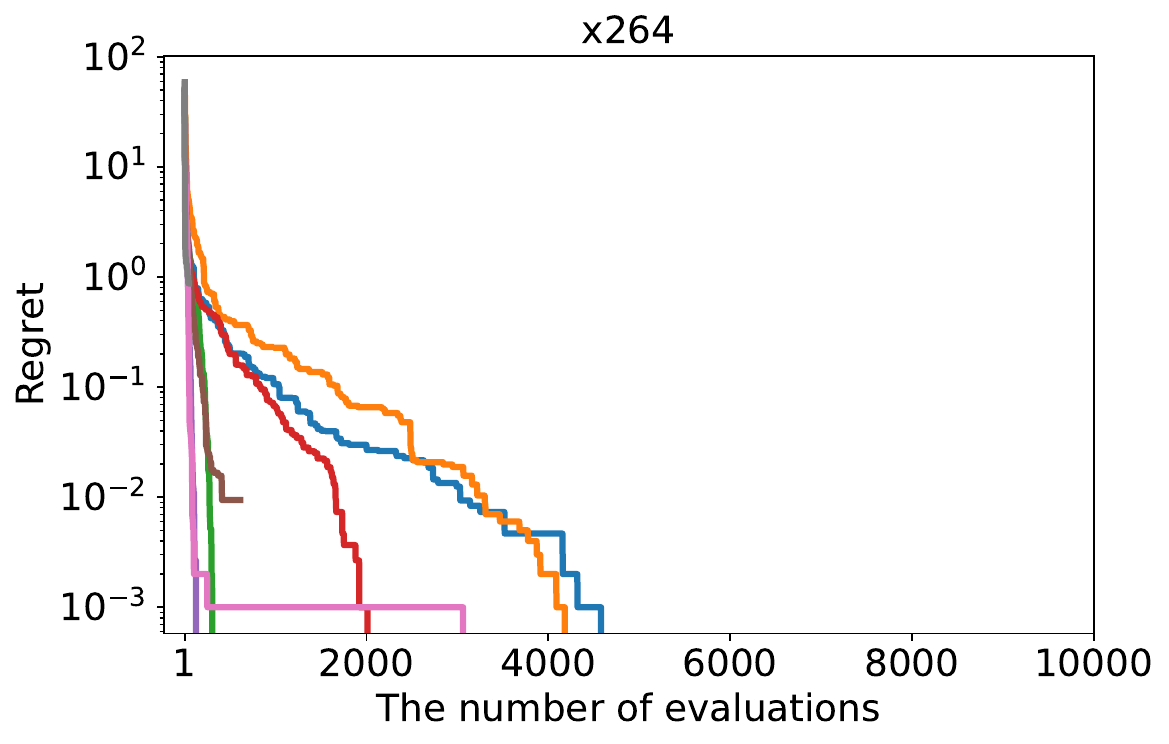}
    \end{minipage}
    \begin{minipage}{0.48\linewidth}
        \includegraphics[width=\linewidth]{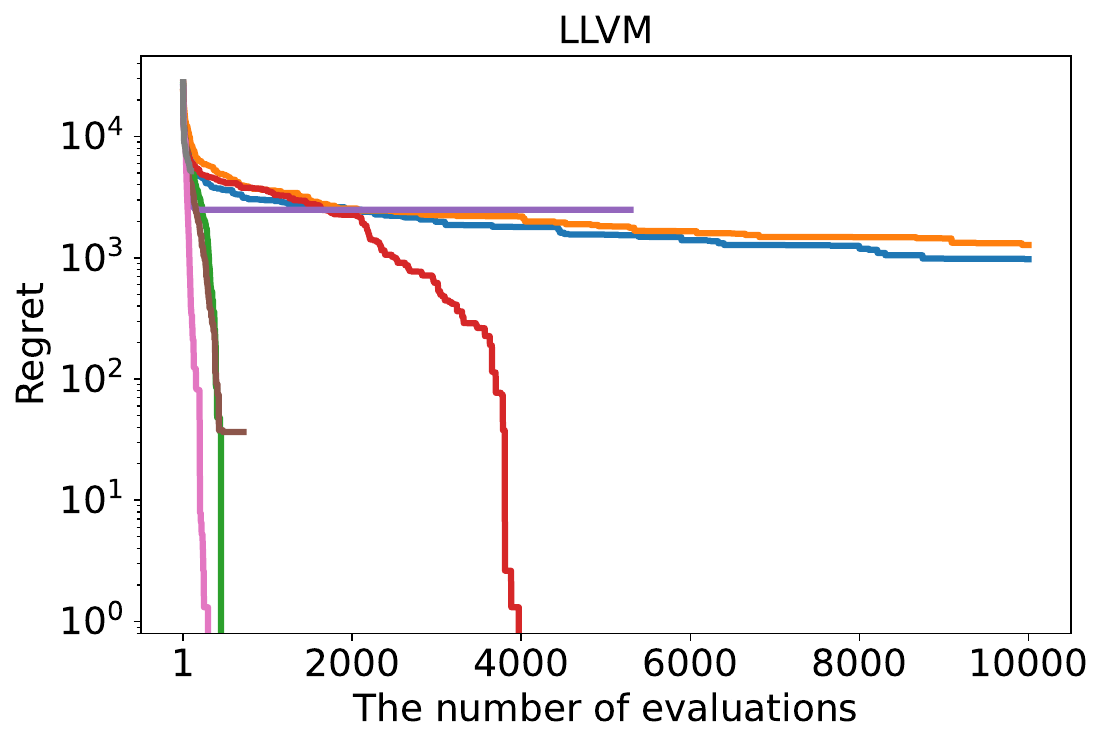}
    \end{minipage}
    

    \begin{minipage}{0.8\linewidth}
        \includegraphics[width=0.95\linewidth]{figures/legend.pdf}
    \end{minipage}

    \captionsetup{skip=4pt}
    \caption{Convergence trajectories of the 8 optimisers on the 4 representative systems (\textsc{BATIK}, \textsc{NGINX}, \textsc{x264}, and \textsc{LLVM}) during the tuning process with a varying budget of up to 10,000 evaluations. Each coloured line represents the mean regret on 30 independent runs.}

    \label{fig:convergence10000}
\end{figure}

Figure~\ref{fig:convergence10000} shows convergence trajectories of the 8 optimisers on the 4 representative systems, i.e., \textsc{BATIK}, \textsc{NGINX}, \textsc{x264}, and \textsc{LLVM}. 
The convergence trajectories on the other 18 systems can be found in the supplementary material. 
As can be seen in Figure~\ref{fig:convergence10000}, FLASH, GA, and IRACE are able to locate the optimal configuration on the four systems, e.g., under 2,000, 3,300, and 4,000 evaluations on \textsc{BATIK}, respectively.

Among the other model-based optimisers, SMAC and TPE show mixed performance. 
They achieve strong results on some systems, such as \textsc{Z3} and \textsc{BROTLI}, but fail on some others, as evidenced by their higher regret values and standard deviations. 
For example, SMAC performs best on \textsc{Z3}, while on \textsc{ExaStencils}, it ranks second to last based on its mean regret. 
Similarly, TPE performs best on \textsc{SQLite}, but on \textsc{Kanzi}, it ranks second to last.

Interestingly, both RS and SA perform well, achieving the best results in 18 out of the 22 systems after 3,000 evaluations. 
This is consistent with the findings of~\citet{bergstra2012random}, who demonstrated the effectiveness of RS in general optimisation problems. 
It also aligns with~\citet{arostegui2006empirical}, which suggested that SA tends to perform well when a larger search budget is available.

\begin{quotebox}
   \noindent
   \textit{\textbf{Finding 4:} Model-free optimisers (e.g., GA and IRACE) can rival or even outperform model-based optimisers when provided with a large budget. }
\end{quotebox}





Based on the above analysis, we can conclude that:

\begin{tcbitemize}[raster columns=1, raster rows=1]
    \tcbitem[myhbox={}{Answer to RQ2}]  
    \textit{Under a sufficient budget (up to 10,000), some model-free optimisers catch up and become the best-performing algorithms, such as GA and IRACE. Model-based optimisers are less competitive, with the exception of FLASH, which achieves the best in 20/22 systems regardless of the budget tightness.}
\end{tcbitemize}

\subsection{How Can the Answers to RQ1 and RQ2 Be Explained?}


\subsubsection{Method}\label{sec:RQ2_method}

In Sections~\ref{sec:RQ1} and ~\ref{sec:RQ2}, we have seen that 
FLASH works very well across different budget levels, achieving the statistically best results on most of the systems.
FLASH is designed to favour configurations close to the current best, which results in a greedy search behaviour. Consequently, it can quickly reach a local optimum. However, it is not clear why this behaviour makes FLASH competitive and, in many cases, enables it to outperform other optimisers. Is this because the considered configurable systems contain very few local optima (i.e., a clear global structure exists), or is there another factor at play? Here, we aim to explore this question by investigating the ``distribution'' of local optima within these systems. 
Specifically, we consider the number of local optima, their quality, basin size, and distance to the global optimum. A system with very few local optima may allow FLASH to perform well. Similarly, a system with high-quality local optima that have large basin sizes (i.e., capable of attracting many nearby configurations) also favours FLASH. In contrast, a system with numerous low-quality local optima may hinder FLASH’s performance.


\begin{algorithm}[!ht]\small
    \DontPrintSemicolon
    \caption{Basin of Attraction Identification}
    \label{alg:basin}
    
    \KwIn {$\mathcal{X}$: Configuration space;
    $f(\cdot)$: Performance measure;
    $\mathcal{N}(x)$: Neighbours of $x$.}

    \ForEach{$x \in \mathcal{X}$}{
        \If{$f(x) < f(x_{nbr}) \ \ \forall x_{nbr} \in \mathcal{N}(x)$}{
            $\mathcal{B}(x) \leftarrow \emptyset$  \tcp*{Initialise each basin of attraction}
        }
    }
    
    \ForEach{$x \in \mathcal{X}$}{
        $x_{\text{curr}} \leftarrow x$ \;
        \While{$\exists x' \in \mathcal{N}(x_{\text{curr}})$ \text{with} $f(x') < f(x_{\text{curr}})$}{
            $x_{\text{curr}} \leftarrow \arg\min_{x' \in \mathcal{N}(x_{\text{curr}})} f(x')$ \tcp*{move to best neighbour}
        }

        $\mathcal{B}(x_{\text{curr}}) \leftarrow\mathcal{B}(x_{\text{curr}}) \cup \{x\}$
    }
    \KwOut {Basins of Attraction $\{\mathcal{B}(x)\}$}
    \vspace{-2pt}
\end{algorithm}

Formally, a configuration $x$ is defined as a local optimum if none of its neighbours yields better performance: 
$f(x) < f(x_{nbr}) \  \forall x_{nbr} \in \mathcal{N}(x),$
where $f(\cdot)$ denotes the objective function and $\mathcal{N}(x)$ is the set of neighbours (configurations) that differ from $x$ in one option (i.e., with Hamming distance of $1$). In addition, each local optimum has an associated basin of attraction, defined as the set of configurations that eventually converge to that optimum under a local search process: 
\begin{equation}
    \mathcal{B}(x) = \{\, x' \in \mathcal{X} \mid \text{local search from}\  x' \text{ terminates at } x \,\},
\end{equation}
where $\mathcal{X}$ is the configuration space. The basin size is given by $|\mathcal{B}(x)|$, i.e., the number of configurations belonging to this basin. 
Specifically, we identify basins using a local search procedure: for every configuration, we iteratively explore its neighbours until no further improvement can be found, and add the configuration to the basin of the corresponding local optimum. 
The detailed procedure is summarised in Algorithm~\ref{alg:basin}. 




\begin{figure*}[!ht]
    \centering

    \begin{subfigure}[b]{\linewidth}
        \begin{minipage}{0.32\linewidth}
            \includegraphics[width=\linewidth]{figures/RQ1/LLVM_500.pdf}
        \end{minipage}\hfill
        \begin{minipage}{0.32\linewidth}
            \includegraphics[width=\linewidth]{figures/RQ2/LLVM_10000.pdf}
        \end{minipage}\hfill
        \begin{minipage}{0.32\linewidth}
            \includegraphics[width=\linewidth]{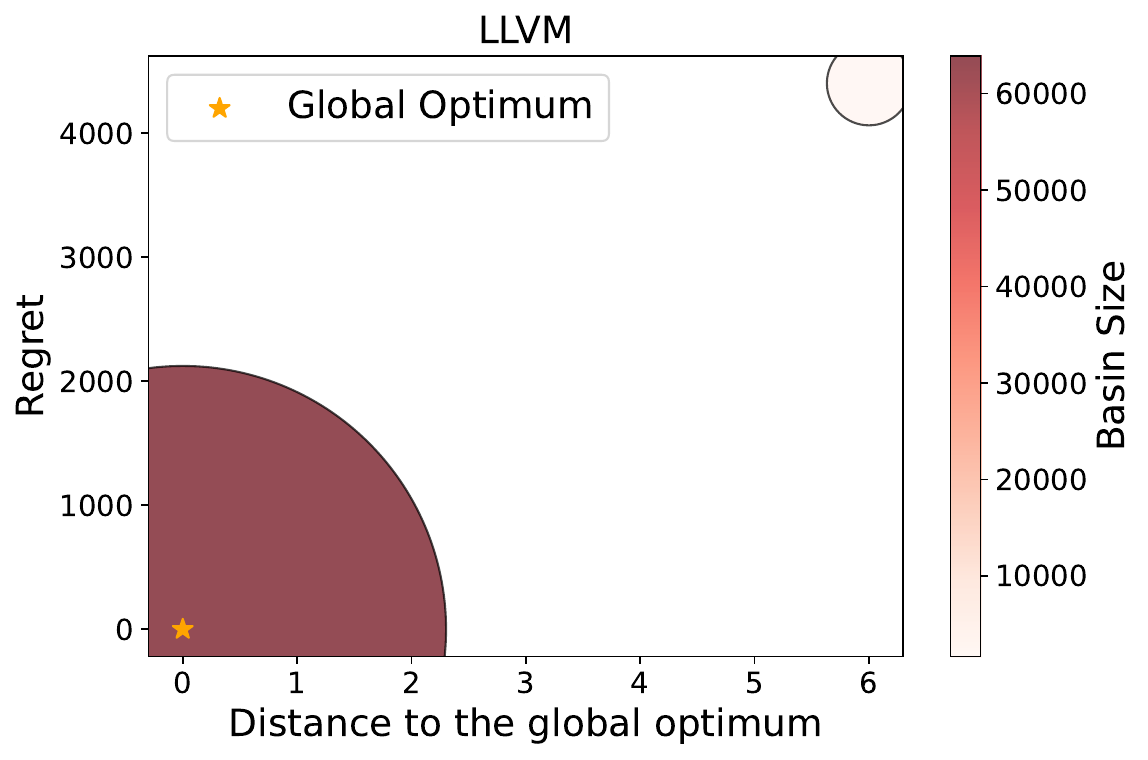}
        \end{minipage}

        \caption{Convergence trajectories and distribution of local optima of \textsc{LLVM}, with a search space size of $6.55 \times 10^4$. 
        }
    \end{subfigure}

    \begin{subfigure}[b]{\linewidth}
        \begin{minipage}{0.32\linewidth}
            \includegraphics[width=\linewidth]{figures/RQ1/x264_500.pdf}
        \end{minipage}\hfill
        \begin{minipage}{0.32\linewidth}
            \includegraphics[width=\linewidth]{figures/RQ2/x264_10000.pdf}
        \end{minipage}\hfill
        \begin{minipage}{0.32\linewidth}
            \includegraphics[width=\linewidth]{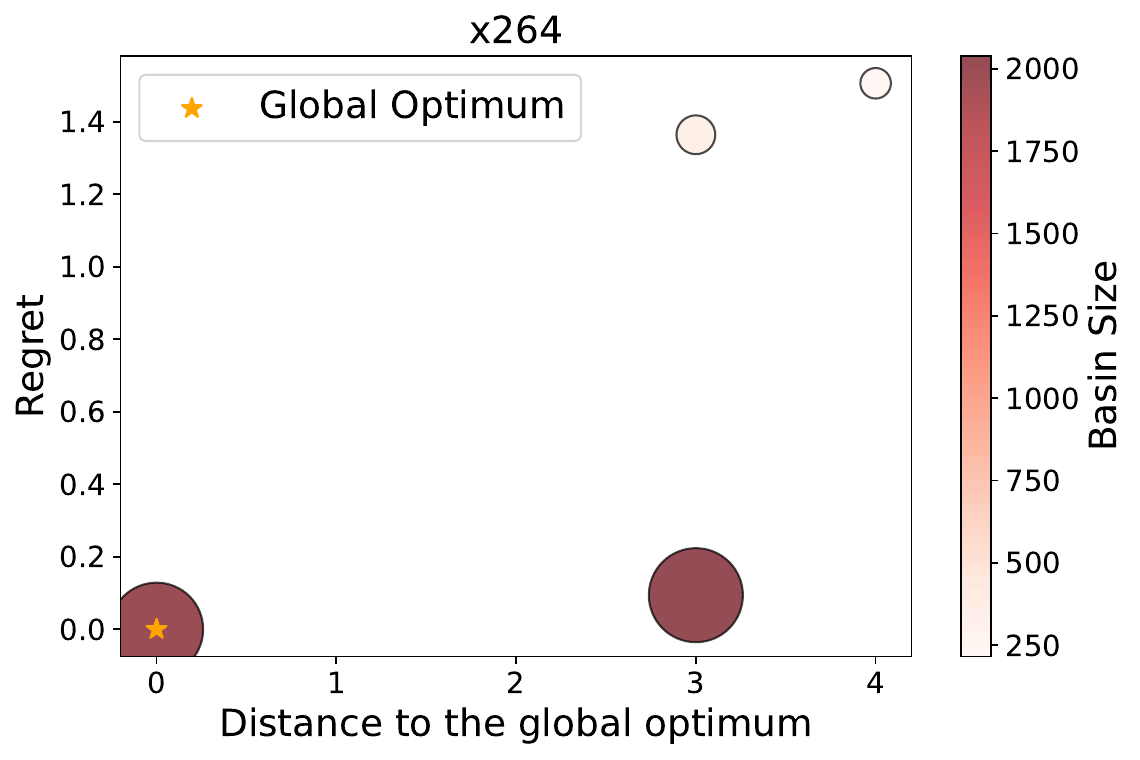}
        \end{minipage}

    \caption{Convergence trajectories and distribution of local optima of \textsc{x264}, with a search space size of $4.60 \times 10^3$. 
    }
    \end{subfigure}

    \begin{subfigure}[b]{\linewidth}
        \begin{minipage}{0.32\linewidth}
            \includegraphics[width=\linewidth]{figures/RQ1/batlik_500_old.pdf}
        \end{minipage}\hfill
        \begin{minipage}{0.32\linewidth}
            \includegraphics[width=\linewidth]{figures/RQ2/batlik_10000.pdf}
        \end{minipage}\hfill
        \begin{minipage}{0.32\linewidth}
            \includegraphics[width=\linewidth]{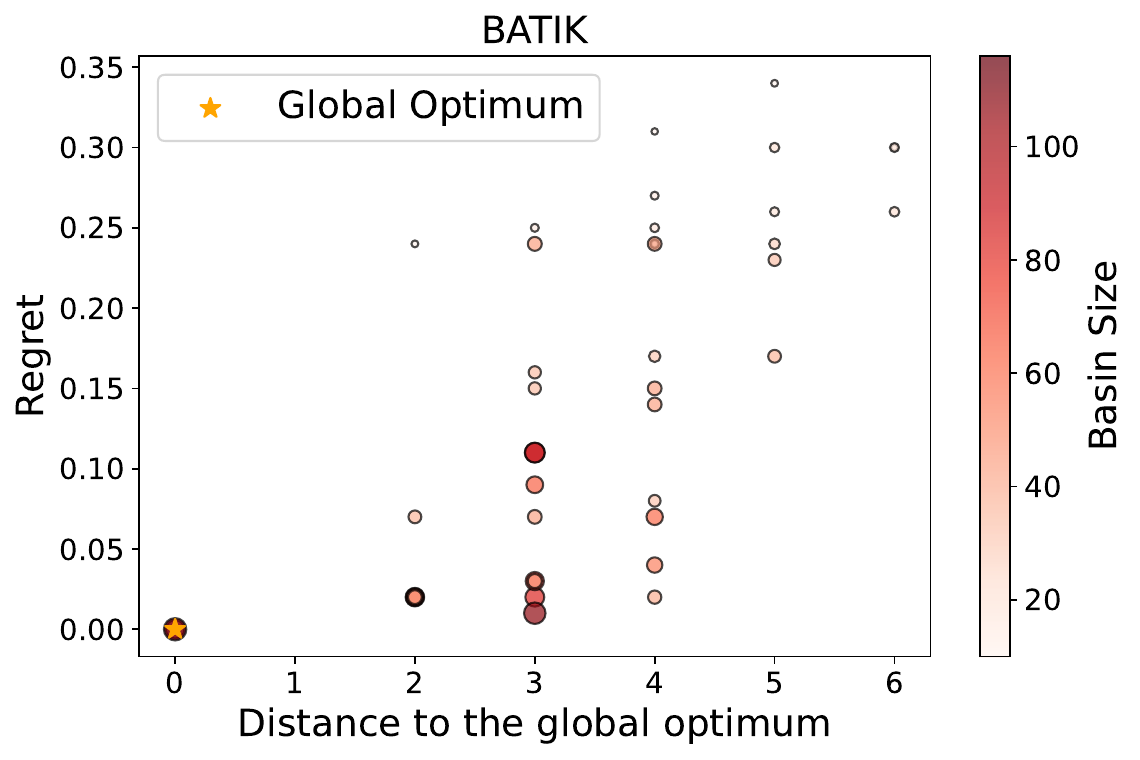}
        \end{minipage}
        \caption{Convergence trajectories and distribution of local optima of \textsc{BATIK}, with a search space size of $1.22 \times 10^4$. 
        }
    \end{subfigure}

    \begin{subfigure}[b]{\linewidth}
        \begin{minipage}{0.32\linewidth}
            \includegraphics[width=\linewidth]{figures/RQ1/nginx_500.pdf}
        \end{minipage}\hfill
        \begin{minipage}{0.32\linewidth}
            \includegraphics[width=\linewidth]{figures/RQ2/nginx_10000.pdf}
        \end{minipage}\hfill
        \begin{minipage}{0.32\linewidth}
            \includegraphics[width=\linewidth]{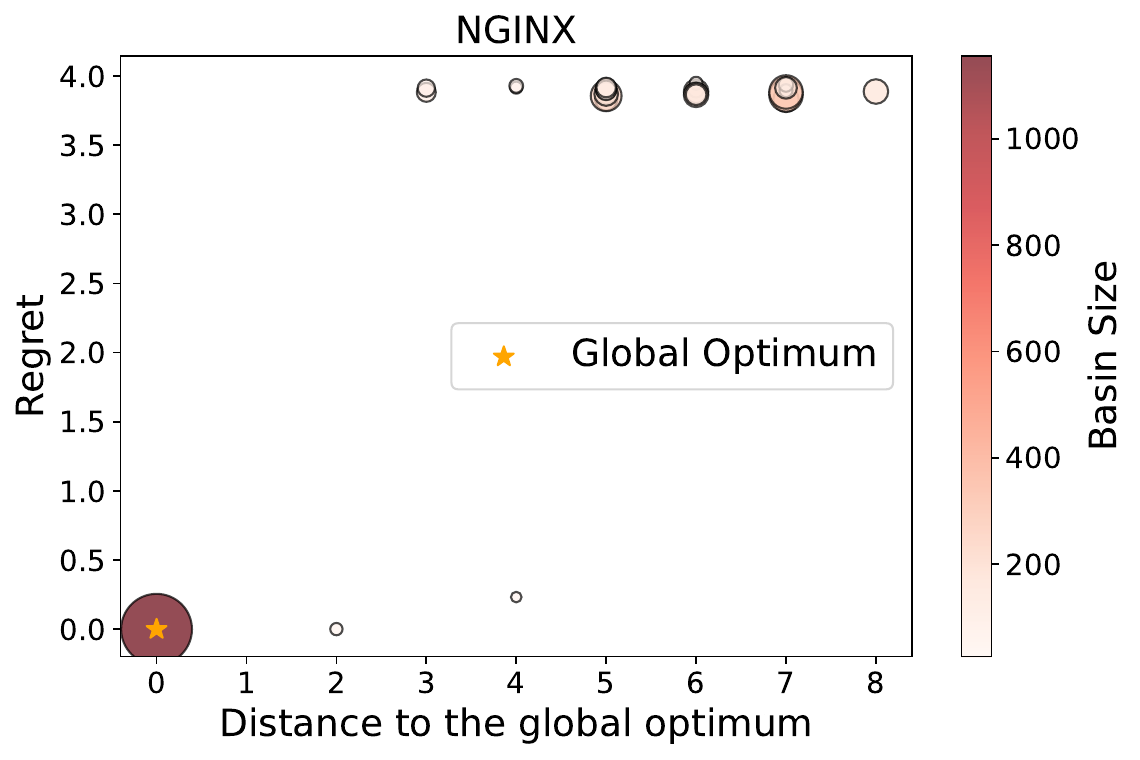}
        \end{minipage}
        
        \begin{minipage}{0.95\linewidth}
            \centering
            \includegraphics[width=0.8\linewidth]{figures/legend.pdf}
        \end{minipage}
        
        \caption{Convergence trajectories and distribution of local optima of \textsc{NGINX}, with a search space size of $1.53 \times 10^4$. 
        }
    \end{subfigure}

    \caption{Distributions of local optima (right plots) of the 4 representative systems (\textsc{BATIK}, \textsc{LLVM}, \textsc{x264}, and \textsc{NGINX}), as well as convergence trajectories of the 8 optimisers on them. 
    The left plots and middle plots show the convergence trajectories under commonly used budget ranges (up to 500 evaluations) and larger budgets (up to 10,000 evaluations), respectively, as previously shown in Figures~\ref{fig:convergence500} and~\ref{fig:convergence10000}. 
    The right plots illustrate the distribution of local optima. 
    Each circle represents a local or global optimum, where the x-axis shows its Hamming distance to the global optimum and the y-axis shows its regret. Both circle radius and colour intensity reflect basin size, where larger and darker circles correspond to larger basins of attraction. It is worth noting that all global optima are highlighted by orange stars, located in the lower-left region of these plots. 
    }
    
    \label{fig:localoptima}
\end{figure*}

\subsubsection{Findings}

Figure~\ref{fig:localoptima} shows convergence trajectories of the 8 optimisers and the distributions of local optima on the 4 representative systems (i.e., \textsc{LLVM}, \textsc{x264}, \textsc{BATIK}, and \textsc{NGINX}), as discussed in Sections~\ref{sec:RQ1} and~\ref{sec:RQ2}. 
The left plots and middle points show the convergence trajectories under commonly used budget ranges (up to 500 evaluations) and larger budgets (up to 10,000 evaluations), respectively, as previously shown in Figures~\ref{fig:convergence500} and~\ref{fig:convergence10000}. 
The right plots illustrate the distribution of local optima, where each circle represents either a local or global optimum. 
For each local optimum, we measure three properties: (i) its quality (i.e., regret) relative to the global optimum (y-axis), (ii) the size of its basin of attraction (represented by both circle radius and colour intensity for clarity), and (iii) its Hamming distance to the global optimum (x-axis). 
It is worth noting that all global optima are highlighted by orange stars, located in the lower-left region of these plots. 
The distributions of local optima on the other 18 systems can be found in the supplementary material.

Firstly, we consider \textsc{LLVM}, which has a relatively large search space ($6.55 \times 10^4$). As shown in Figure~\ref{fig:localoptima}(a), FLASH achieves the best performance across all budget levels. The reason is straightforward: the landscape is rather unimodal (i.e., it contains very few local optima). A majority of configurations are attracted to the global optimum, which has a large basin of attraction (see the lower-left circle with an orange star). There exists only one low-quality local optimum with a small basin. As a result, despite the relatively large configuration space, FLASH is able to find the global optimal configuration quickly due to its greedy search behaviour. 

\begin{quotebox}
   \noindent
   \textit{\textbf{Finding 5:} For systems with rather unimodal landscapes, such as \textsc{LLVM}, FLASH with greedy search behaviour consistently delivers the best tuning quality across different budget levels, even when the configuration space is relatively large. 
   }
\end{quotebox}

Next, we turn to \textsc{x264}, which has a search space of $4.60 \times 10^3$. As illustrated in Figure~\ref{fig:localoptima}(b), FLASH performs best when the evaluation budget is limited (e.g., 100 evaluations). After that, it maintains competitive results with a very small regret of $10^{-3}$. 
One possible reason is that, along with the global optimum, there exists a high-quality local optimum with a large basin (around 2,000 in size). This allows FLASH to efficiently identify a good configuration, though sometimes it may require more evaluations to move beyond that local optimum in later stages. 
In addition, for such landscapes, exploration-oriented optimisers such as TPE and GA can efficiently find the optimal configuration under fairly tight budgets (120 evaluations for TPE and 300 evaluations for GA, respectively). This may be due to their exploratory ability, enabling them to search for the basin of the global optimum.

\begin{quotebox}
   \noindent
   \textit{\textbf{Finding 6:} In systems having high-quality local optima with large basins, such as \textsc{x264}, FLASH can rapidly obtain high-quality configurations under very tight budgets.
   }
\end{quotebox}

Now we move to \textsc{BATIK} and \textsc{NGINX}. It is worth noting that \textsc{BATIK} and \textsc{NGINX} share similar attributes: both have a dimensionality of 11 and similar search space sizes ($1.22 \times 10^4$ for \textsc{BATIK} and $1.53 \times 10^4$ for \textsc{NGINX}). However, FLASH and GA perform well on \textsc{BATIK} yet not well on \textsc{NGINX}.

\begin{quotebox}
   \noindent
   \textit{\textbf{Finding 7:} Systems with similar dimensionality and search space size, such as \textsc{BATIK} and \textsc{NGINX}, optimisers can exhibit very different tuning behaviours, suggesting that tuning difficulty depends more on the structure of the landscape than on its attributes.}
\end{quotebox}

Regarding \textsc{BATIK}, as shown in Figure~\ref{fig:localoptima}(c), FLASH achieves the best performance regardless of the budget levels, though there are many local optima (many of which are of low quality). 
This occurrence may be attributed to the fact that although there are lots of low-quality local optima, they have small basin sizes (many smaller than five). At the same time, there are several high-quality local optima with large basin sizes (around 100). 
Consequently, FLASH is very likely to rapidly hit the high-quality configurations, and even if the search moves into a low-quality local optimum, the small basin size allows it to escape readily. 
This suggests that even in systems with many local optima, FLASH can still achieve favourable tuning quality as long as its low-quality basins remain small, compared to high-quality basins.

\begin{quotebox}
   \noindent
   \textit{\textbf{Finding 8:} For systems with many local optima, provided high-quality ones having significantly larger basin sizes than low-quality ones (e.g., \textsc{BATIK}), FLASH is able to efficiently obtain high-quality configurations regardless of the budget level. 
   }
\end{quotebox}

As for \textsc{NGINX} (Figure~\ref{fig:localoptima}(d)), FLASH performs well only under a very limited budget (e.g., 50 evaluations). After that, its performance stagnates for a considerable period, i.e., with no significant improvement in tuning quality, before finally converging to the global optimum when the evaluation budget reaches approximately 800. One reason for this phenomenon lies in the structural characteristics of NGINX’s search space: it contains a large number of local optima, most of which are of low quality (with fairly large basin sizes).
Furthermore, these local optima lie far from the global optimum, with their Hamming distances to the global optimum ranging from 3 to 8. 
Hence, in the early stages of the tuning process (with a small budget), FLASH is likely to be trapped in low-quality local optima, leading to prolonged performance stagnation. 
Only when the evaluation budget increases to a fair amount (e.g.,~800) can FLASH accumulate enough samples to escape the local optima, traverse a broader region of the configuration space, and identify the global optimal configuration. 
In contrast, SMAC performs best under a fairly tight budget (e.g., 300 evaluations), and TPE excels when the evaluation budget exceeds 400, though it also demonstrates strong tuning quality earlier. This can be attributed to their exploratory search strategies that allow them to escape the local optima. Furthermore, as the evaluation budget increases (e.g., 600 evaluations), GA becomes one of the best optimisers due to its global search ability.

\begin{quotebox}
   \noindent
   \textit{\textbf{Finding 9:} For systems having many low-quality local optima with fairly large basin sizes, such as \textsc{NGINX}, FLASH may be stuck at the beginning and need more evaluations to locate the global optimum. In contrast, TPE becomes more effective, and also GA shows strong performance after a while.}
\end{quotebox}




Based on the above analysis, we can conclude that:

\begin{tcbitemize}[raster columns=1, raster rows=1]
    \tcbitem[myhbox={}{Answer to RQ3}]  
    \textit{The same optimiser exhibits different tuning quality mainly due to both budget tightness and the configuration landscape of systems (e.g., the quality of local optima and the size of their basins).
    Specifically, FLASH generally performs well regardless of budget tightness -- except in cases where a system contains numerous low-quality local optima with relatively large basin sizes. 
    In addition, exploration-oriented optimisers (e.g., GA) are more effective in maintaining exploration and escaping local optima, enabling them to eventually converge to the global optimum when more evaluations are available.
    }
\end{tcbitemize}

\section{Implications}\label{sec:Implications}

Based on the above observations and analysis, we now provide actionable recommendations for industry practitioners and configuration tuning researchers dealing with tuning problems under a specific budget.

\subsection{For Industry Practitioners}

When tackling a software configuration tuning problem within a given budget, practitioners may refer to the following guidelines.

\begin{itemize}
    \item With a normally tight budget (e.g., less than 500 evaluations), use FLASH (\textbf{\textit{Findings 1 and 3}}).


    \item When more than a normally tight budget is available (e.g., more than 500 evaluations), use GA (\textbf{\textit{Finding 4}}).

    \item When the budget is very generous (e.g., more than 4000 evaluations), try either GA or IRACE (\textbf{\textit{Finding 4}}).
    
\end{itemize}

Note that the above recommendations are based on the situation that the system is purely black-box, i.e., no other knowledge is available. Sometimes, there is some knowledge about a given system available \textit{a priori}, e.g., ruggedness of landscape (search space). 
For example, the search space of \textsc{Storm} has been known to be fairly rugged~\cite{chen2024mmo,jamshidi2016uncertainty}. 
In this case, try to factor such knowledge in when choosing an optimiser. 

\begin{itemize}
    \item For systems with rather unimodal landscapes, regardless of the budget level, use FLASH (\textbf{\textit{Finding 5}}).

    \item For systems with small search spaces but large basins around high-quality local optima (e.g., thousands of configurations), regardless of the budget level, use FLASH (\textbf{\textit{Finding 6}}).

    \item For systems with small search spaces (e.g., ten thousand) and basins (e.g., dozens), regardless of the budget level, use FLASH (\textbf{\textit{Finding 8}}).

    \item For systems with large basins associated with low-quality local optima, use FLASH under a rather limited budget (e.g., 100 evaluations), use TPE under a fairly limited budget (e.g., 100-500 evaluations), and use both TPE and GA under a more generous budget (e.g., 500 evaluations) (\textbf{\textit{Finding 9}}).

    
\end{itemize}



\subsection{For Configuration Tuning Researchers}
For configuration tuning researchers building new optimisers, our study provides several design recommendations:

\begin{itemize}

    \item \textbf{Budgets matter:} We observe that optimisers can behave quite differently when the budget is small versus when it is large. Exploitation-oriented optimisers (e.g., FLASH) tend to excel under limited budgets, while exploration-oriented optimisers (e.g., GA) become more effective with larger budgets. This suggests that future optimisers could adapt such strategies to the available budget, e.g., focusing more on exploitation when resources are tight and shifting towards exploration as the budget grows (\textbf{\textit{Findings 1-4}}).

    \item \textbf{Simplicity can be powerful:} Our results show that FLASH, which uses a simple model and a greedy strategy, still works very well across many systems. This means that future optimisers may not always have to be complex — simple and efficient designs can also be good choices (\textbf{\textit{Answer to \textbf{RQ2}}}). 
    
    \item \textbf{Landscape matters:} Our study shows that tuning quality depends less on system attributes (e.g., the number of options) and more on the structure of the configuration landscape, such as how many local optima there are, how good they are, and how big their basins are. This suggests that future optimisers could try to estimate these patterns (e.g., by sampling) and adjust their behaviour accordingly, such as focusing locally when high-quality optima with large basins are present and vice versa (\textbf{\textit{Findings 5-9}}). 
    

\end{itemize}

\section{Threats to Validity}\label{sec:threats}

As with any empirical study, certain biases could influence the final results. Therefore, the conclusions drawn from this work should be interpreted with the following limitations in mind:

\begin{itemize}

    \item \textbf{System bias:} This study investigates the 22 configurable systems, spanning diverse real-world software domains, programming languages, dimensionality, and variable types.
    However, the findings may not generalise to all types of configurable systems.

    \item \textbf{Optimiser bias:} \re{We have chosen the 8 model-based and model-free optimisers according to the criteria described in Section~\ref{sec:method}. The conclusions may not generalise to other important optimisers, SPL optimisers~\cite{oh2023finding,hierons2016sip,oh2017finding,guegain2023configuration}.
    
    }

    
    


    \item \textbf{Parameter bias:} Each optimiser is applied using the default parameter settings recommended in the original papers. While this ensures consistency, it is possible that alternative parameter settings could yield different performance patterns, which may affect the conclusions.
    

    \item \textbf{Measurement bias:} While measurement bias may exist, the datasets we used mitigate this issue by using repeated measurements for each configuration, with the median or mean taken as the final performance value~\cite{muhlbauer2023analysing,krishna2020whence,weber2023twins}.
    

    
    \item \textbf{Landscape analysis bias:} To understand why FLASH with greedy search behaviour performs consistently well, we consider the distribution of local optima. However, other aspects of the systems' landscape may also affect the performance of optimisers, such as the ruggedness level of the search space~\cite{huang2025rethinking}.


\end{itemize}



\section{Related Work}\label{sec:related_work}

Over the past decades, lots of studies have investigated the tuning quality of various tuning methods for configurable systems~\cite{li2025csat,chen2023performance,cao2024etune,cao2024ptssbench,ha2022uncertainty,lee2024k2vtune,chen2025accuracy}. 
Many of these studies focused on developing new tuning methods to enhance tuning quality~\cite{lee2024k2vtune,ha2022uncertainty,yang2024vdtuner,zhan2024knobtune} (e.g., LATuner for database systems~\cite{fan2024latuner}), though only general-purpose optimisers across diverse configurable systems are in the scope of this study, as stated in Section~\ref{sec:method}. 

In addition, some studies conducted empirical evaluations, comparing existing tuning methods in terms of specific purposes~\cite{chen2025accuracy,chen2023performance,zhang2021evolutionary}. 
Additionally, some studies use fitness landscape analysis for configurable systems~\cite{lustosa2024learning,huang2025rethinking}.
We detail the above studies in the following. 

\subsection{Empirical Studies on Configuration Tuning}


A number of empirical studies have been conducted on configurable systems with different purposes~\cite{van2021inquiry,li2024large,chen2025accuracy,chen2023performance,zhang2021evolutionary,xu2015hey,han2016empirical}. 
Among them, some work on model-free tuning methods~\cite{liao2022empirical,chen2023performance}. 
For example, \citet{chen2023performance} evaluate the role of performance requirements in tuning objectives and identify when they are advantageous or detrimental. 
Additionally, some focus on model-based tuning methods~\cite{van2021inquiry,zhang2021facilitating}. 
~\citet{van2021inquiry} show that model-based tuning methods can find configurations superior to those of human experts, achieving up to 45\% better performance. 
Furthermore, some studies work on both model-based and model-free tuning methods~\cite{chen2025accuracy}. 
For example, \citet{chen2025accuracy} challenge the common but unverified belief in model-based configuration tuning that higher model accuracy leads to better tuning results.


However, these efforts focus on different aspects (e.g., the role of performance aspirations~\cite{chen2023performance}), and therefore fix their budgets at specific (tight) levels. In contrast, our study aims to systematically investigate the performance of tuning methods under varying budgets, considering a much broader range (up to 10,000 evaluations) and 22 configurable systems spanning diverse domains, programming languages, dimensionalities, and variable types.

\subsection{Fitness Landscape Analysis for Configuration Tuning}

Among the studies that use fitness landscape analysis for configuration tuning~\cite{lustosa2024learning,huang2025rethinking,jamshidi2017transfer,chen2022planning,chen2025accuracy}, one way is to integrate it directly into the tuning process~\cite{lustosa2024learning}. For example, \citet{lustosa2024learning} employ fitness landscape analysis to guide tuning in software project health prediction, where their method clusters the search space to identify informative regions and then selectively explores representative configurations while jointly optimising multiple objectives. 
In addition, some studies focus on understanding or extracting the characteristics of performance distribution~\cite{huang2025rethinking,jamshidi2017transfer}. 
For example,~\citet{jamshidi2017transfer} propose a framework to understand the intricate relationship between software configurations and system performance. 
However, no prior work has explored the relationship between tuning quality and budget levels using fitness landscape analysis.

\section{Conclusion}\label{sec:Conclusion}

This study systematically evaluated eight well-established optimisers across 22 configurable software systems under varying budgets up to 10,000 evaluations. 
Our findings confirm that model-based optimisers (e.g., SMAC) excel under limited budgets, while model-free optimisers (e.g., GA and IRACE) become competitive as the budget increases.



In addition, interestingly, we also found that FLASH consistently performs very well on most systems regardless of the budget. One important reason behind this is that many configurable software systems possess reasonably good local optima, which also have fairly large basins of attraction.
This allows optimisers with greedy search behaviour (e.g., FLASH) to achieve strong performance since they can quickly 
``hit'' a high-quality local optimal configuration.
Our work also shows a way of explaining observed empirical results through fitness landscape analysis, indicating the usefulness of investigating the correlation between tuning quality of optimisers and systems' landscape. In this regard, some recent work (e.g., \cite{huang2025rethinking}) may be beneficial for further understanding configurable systems.

\bibliographystyle{ACM-Reference-Format}
\bibliography{sample-base}










\end{document}